\documentclass{article}
\usepackage[utf8]{inputenc}
\usepackage{amsmath} 
\usepackage{amssymb}
\usepackage{graphicx}
\usepackage{subfig}
\usepackage{appendix}
\usepackage[svgnames]{xcolor}
\usepackage{soul}
\usepackage[stable]{footmisc}
\usepackage{caption}
\usepackage{authblk}

\title{Constraints on the energy spectrum of the diffuse cosmic neutrino flux from the ANTARES neutrino telescope}
\date{\today}

\usepackage{hyperref}

\hypersetup{
  colorlinks=true,        % false: boxed links; true: colored links
  linkcolor=Violet,          % color of internal links formule
  citecolor=Violet,        % color of links to bibliography citazioni
  filecolor=cyan,      % color of file links listato
  urlcolor=Teal          % color of external links link
}

\begin{document}

\author[1,2]{A.~Albert}
\author[3]{S.~Alves}
\author[4]{M.~Andr\'e}
\author[5]{M.~Ardid}
\author[5]{S.~Ardid}
\author[6]{J.-J.~Aubert}
\author[7]{J.~Aublin}
\author[7]{B.~Baret}
\author[8]{S.~Basa}
\author[7]{Y.~Becherini}
% ...

\affil[1]{\scriptsize{Universit\'e de Strasbourg, CNRS,  IPHC UMR 7178, F-67000 Strasbourg, France}}
\affil[2]{\scriptsize Universit\'e de Haute Alsace, F-68100 Mulhouse, France}
\affil[3]{\scriptsize{IFIC - Instituto de F\'isica Corpuscular (CSIC - Universitat de Val\`encia) c/ Catedr\'atico Jos\'e Beltr\'an, 2 E-46980 Paterna, Valencia, Spain}}
\affil[4]{\scriptsize{Technical University of Catalonia, Laboratory of Applied Bioacoustics, Rambla Exposici\'o, 08800 Vilanova i la Geltr\'u, Barcelona, Spain}}
\affil[5]{\scriptsize{Institut d'Investigaci\'o per a la Gesti\'o Integrada de les Zones Costaneres (IGIC) - Universitat Polit\`ecnica de Val\`encia. C/  Paranimf 1, 46730 Gandia, Spain}}
\affil[6]{\scriptsize{Aix Marseille Univ, CNRS/IN2P3, CPPM, Marseille, France}}
\affil[7]{\scriptsize{Universit\'e Paris Cit\'e, CNRS, Astroparticule et Cosmologie, F-75013 Paris, France}}
\affil[8]{\scriptsize{Aix Marseille Univ, CNRS, CNES, LAM, Marseille, France }}
% ...

\maketitle

{\hypersetup{linkcolor=DodgerBlue}
  % or \hypersetup{linkcolor=black}, if the colorlinks=true option of hyperref is used
  \tableofcontents
}

\abstract{High-significance evidences of the existence of a high-energy diffuse flux of cosmic neutrinos have emerged in the last decade from several observations by the IceCube Collaboration. The ANTARES neutrino telescope took data for 15 years in the Mediterranean Sea, from 2007 to 2022, and collected a high-purity all-flavour neutrino sample. The search for a diffuse cosmic neutrino signal using this dataset is presented in this article. This final analysis did not provide a statistically significant observation of the cosmic diffuse flux. However, this is converted into limits on the properties of the cosmic neutrino spectrum. In particular, given the sensitivity of the ANTARES neutrino telescope between 1 and 50~TeV, constraints on single-power-law hypotheses are derived for the cosmic diffuse flux below 20~TeV, especially for power-law fits of the IceCube data with spectral index softer than 2.8.}

\section{Introduction}\label{sec:intro}

A major goal pursued with neutrino telescopes is the detection of high-energy neutrinos of cosmic origin in the TeV~--~PeV energy range. Neutrino telescopes are three-dimensional arrays of photodetectors where the arrival time and the deposited charge of Cherenkov photons induced by relativistic charged particles in a transparent medium, such as water or ice, is measured. If these charged particles arise from the interaction of a neutrino, its properties can be determined from the detected photon patterns~\cite{bib:markov, bib:spiering}.

Two main event topologies can be observed in neutrino telescopes: \textit{tracks}, produced by the long-lived and penetrating muons induced by charged current $\nu_\mu$ weak interactions; \textit{showers}, produced by electromagnetic and hadronic cascades coming out of the interaction vertex in all-flavour neutrino weak interactions (both charged and neutral current). Since muons can travel several kilometres before being observed, their tracks can be detected from neutrino interactions occurring in a very large volume surrounding the instrumented volume. Showers are instead much more compact (a few-metres long), and thus can only be observed in the proximity of the detector. Directional reconstruction is optimal for tracks, due to the length of the muon path through the detector, while more difficult for showers which are more compact. The opposite is true for the energy reconstruction: most of the light from the neutrino interaction products can be observed in the case of showers, but only part of it in the case of muons coming from interactions far away from the instrumented volume~\cite{bib:spiering}. The word neutrino, here and in the following, will refer to both neutrinos and anti-neutrinos, as neutrino telescopes cannot discriminate between the two.

High-energy cosmic neutrinos can be produced in the aftermath of the interactions of cosmic ray protons and nuclei with matter or radiation fields. Charged pions are the most abundant products of these interactions, and a flux of neutrinos will stem from their decay chains. The energy spectrum of the resulting neutrinos will generally follow that of the primary cosmic ray population: assuming standard scenarios for the acceleration of cosmic rays~\cite{bib:fermiI, bib:fermiII, bib:cr_acc}, their energy spectrum will behave as a power law $dN_p/dE_p \propto E_p^{-\gamma_p}$, with a primary spectral index $\gamma_p$ between 2.0 and 2.4. As such, high-energy neutrinos are probes for primary cosmic rays interacting close to their sources, or along their path in the Universe.

A high-energy diffuse flux of cosmic neutrinos may originate from the ensemble of unresolved individual neutrino sources in the Universe. Because of neutrino oscillations over cosmic distances, equipartition between the three neutrino flavours can be assumed at Earth if neutrinos originate from the decays of pions coming from the interactions of cosmic ray protons~\cite{bib:nuosci_astro}.% Isospin conservation in strong proton-proton interactions also guarantees that an equal amount of positively-charged and negatively-charged pions is produced, thus leading to equal fluxes of neutrinos and anti-neutrinos\footnote{With a small asymmetry due to the fact that positively charged cosmic rays interact.}.

The diffuse cosmic neutrino signal will appear as an excess of high-energy events with respect to events of terrestrial origin --- namely, atmospheric muons and neutrinos produced by cosmic rays interacting in the atmosphere. The energy spectrum of this high-energy neutrino excess is usually modelled as a single unbroken power law for one flavour ($1f$)
\begin{equation}
  \frac{\Phi_{\textrm{astro}}^{1f}(E_\nu)}{C_0} = \phi_{\textrm{astro}} \times \left(\frac{E_\nu}{E_0} \right)^{-\gamma}
  \label{eq:powerlaw}
\end{equation}
with normalisation $\phi_{\textrm{astro}}$ and spectral index $\gamma$. The normalisation constant in equation~\ref{eq:powerlaw} is set to $C_0=10^{-18}\textrm{ (GeV cm$^2$ s sr)$^{-1}$}$ in the rest of this article, with a pivot energy $E_0~=~100\textrm{~TeV}$.

The IceCube Collaboration~\cite{bib:ic} has measured the properties of the high-energy cosmic diffuse flux in several searches~\cite{bib:ic_hese_latest, bib:ic_tracks, bib:ic_cascades, bib:ic_comb}. These analyses provided various estimations for the above parameters describing the diffuse cosmic neutrino spectrum. The Baikal-GVD Collaboration has also reported a mildly-significant observation of the diffuse cosmic neutrino flux in their neutrino data~\cite{bib:gvd_diffuse}. Table~\ref{tab:spectra} summarises the current status of these analyses.

Each IceCube and Baikal-GVD sample has some peculiarities which may play a role in the outcome of the power-law fit: the IceCube High Energy Starting Events sample (HESE)~\cite{bib:ic_hese_latest} is dominated by electron neutrino interactions above 60~TeV, whereas the IceCube track sample~\cite{bib:ic_tracks} collects through-going muons from cosmic $\nu_\mu$ from the Northern Sky in the 15~TeV~--~5 PeV range; the IceCube cascade sample~\cite{bib:ic_cascades} comprises mostly electron neutrinos from the whole sky, but extending to the 10 TeV energy range. The IceCube combined fit~\cite{bib:ic_comb} merges events from different samples together to provide a global fit of the signal. The Baikal-GVD sample~\cite{bib:gvd_diffuse} mostly contains events from electron neutrino interactions above 100 TeV from the Southern Sky. While the measurements are all compatible within their uncertainties, subtle differences are present. These could be attributed to several reasons: the different energy range covered by each analysis; the different flavour composition of the observed signal; the presence of the Galactic Plane in the Southern Sky~\cite{bib:antares_gal, bib:ic_galplane}. In addition, all these results rely on the single unbroken power-law hypothesis, which may not hold at the lowest or highest energies.

\begin{table}
  \centering
  \begin{tabular}{c | c | c}
    Analysis sample & $\phi_{\textrm{astro}}$ & $\gamma$\\
    \hline
    & & \\
    IceCube HESE~\cite{bib:ic_hese_latest} &$2.12^{+0.49}_{-0.54}$ & $2.87^{+0.20}_{-0.19}$\\
    IceCube tracks~\cite{bib:ic_tracks} & $1.44^{+0.25}_{-0.26}$ &  $2.37\pm0.09$\\
    IceCube cascades~\cite{bib:ic_cascades} & $1.66^{+0.25}_{-0.27}$ & $2.53\pm0.07$\\
    IceCube combined~\cite{bib:ic_comb} & $1.80^{+0.13}_{-0.16}$& $2.52\pm0.04$\\
    & &\\
    Baikal-GVD~\cite{bib:gvd_diffuse} & $3.04^{+1.52}_{-1.21}$ & $2.58^{+0.27}_{-0.33}$
  \end{tabular}
  
  \caption{Summary of the results obtained in the search for a diffuse flux of high-energy cosmic neutrinos in the IceCube~\cite{bib:ic_hese_latest, bib:ic_tracks, bib:ic_cascades, bib:ic_comb} and Baikal-GVD~\cite{bib:gvd_diffuse} data, assuming a single unbroken power-law spectrum and following the definition of the normalisation $\phi_{\textrm{astro}}$ and spectral index $\gamma$ given in equation~\ref{eq:powerlaw}. The uncertainties reported in the table refer to the 68\% confidence level intervals.}
  \label{tab:spectra}
\end{table}

The ANTARES neutrino telescope~\cite{bib:antares} provides a complementary view to that of IceCube. Its efficiency for the detection of neutrinos in the 10~--~50~TeV energy range arising from the Southern Sky is similar to that of IceCube even though ANTARES is much smaller in volume. Indeed, background rates in ANTARES are lower than in IceCube, since ANTARES is located at a larger average depth than IceCube, with the uppermost detector elements at a depth of about 2000 and 1400~m, respectively. In the ANTARES case, the selection of a pure neutrino sample does not require using parts of the instrumented volume as a veto as is needed for the HESE sample.

This article describes the search for the cosmic diffuse neutrino flux and the estimation of its properties using the full and final 15-year ANTARES neutrino data sample. The document is organised as follows: section~\ref{sec:selection} describes the detector and the neutrino samples used in the analysis; section~\ref{sec:stat} covers the statistical methods used in this work; the results of the search for the diffuse cosmic neutrino signal are presented in sections~\ref{sec:results} and~\ref{sec:he_events}. Conclusions and outlooks are given in section~\ref{sec:concl}.

\section{The ANTARES detector}\label{sec:selection}

The ANTARES neutrino telescope~\cite{bib:antares} started taking data in February 2007, and was disconnected in February 2022. It was operated continuously during this period, and was the largest underwater neutrino telescope in the world for most of this time. The apparatus was located in the Mediterranean Sea, 40 km off-shore Toulon, France, anchored to the seabed at a depth of 2475 m. It was made of twelve 350-m long mooring lines, each holding 25 triplets of optical modules, pressure-resistant glass spheres each housing a 10-inch photomultiplier tube~\cite{bib:antares_om}. Cherenkov photons were detected by the optical modules, and all collected signals (``hits'') were sent to a shore station where data were processed, filtered, written to file, and then sent to storage for off-line analyses~\cite{bib:antares_daq}. Results have been published by the ANTARES Collaboration on multiple subjects in the search for cosmic neutrinos~\cite{bib:antares_gal, bib:antares_diffuse, bib:antares_ps, bib:antares_atmo, bib:antares_followup, bib:antares_diffuse_icrc}. The decommissioning of the detector ended in May 2022, when the detector elements were removed from the deep sea.

\subsubsection*{Event reconstruction} \label{sec:ereco}

Maximum-likelihood algorithms are employed to determine the properties of the events (direction and energy). For tracks, a multi-step procedure is followed~\cite{bib:antares_ps}: three preliminary fits are executed (a linear $\chi^2$-fit, a least-squares linear fit ``M-estimator'' minimisation, and a simplified likelihood fit); their output is then the starting point of the final maximum-likelihood fit. The probability density function of the hit time residuals, defined as the time differences between the observed and expected hits on the optical modules, are used to determine the incoming track direction. The energy of the muon is estimated by computing the energy deposit per unit track length~\cite{bib:antares_atmo_tr} on the basis of the observed deposited charge on the PMTs, the track length in the instrumented volume, and accounting for the photon detection efficiency of each PMT. In the case of showers, the event reconstruction is divided into four steps~\cite{bib:antares_tantra}: an initial selection of the hits in the detector to remove spurious signals; a least-squares linear fit (``M-estimator'') to preliminarily determine the interaction vertex; a subsequent hit selection based on the result of this position fit; a likelihood maximisation procedure based on probability density functions for the occurrence of hits in time and space given a certain neutrino direction and energy.

Energy reconstruction for showers is optimal given the almost-calorimetric measurement of their signal in the instrumented volume. On the other hand, the energy estimation for tracks lacks such precision. For this reason, the track energy estimation procedure has been recently re-assessed, including detector-dependent effects such as the measured decrease of the optical module efficiency with time~\cite{bib:antaresk40}, and a re-calibration of the energy estimation based on Monte Carlo simulations~\cite{bib:antares_mc}. The achieved energy resolution for tracks is of the order of 0.5 in the logarithm of the muon energy~\cite{bib:antares_atmo_tr}, while for showers it reaches values as low as 10--15\% for electron neutrinos undergoing charged current interactions~\cite{bib:antares_tantra}. The median angular resolution for $\nu_\mu$-induced tracks is around 0.4$^\circ$ at 100~TeV, and around 2$^\circ$ for $\nu_e$-induced showers.

\subsection{Event selection} \label{sec:esel}

The large majority of events detected in neutrino telescopes are \textit{atmospheric muons} produced in the extensive air showers emerging from cosmic rays interactions. This overwhelming background can be removed by selecting only those events that have crossed the Earth --- coming from below the horizon --- since only neutrinos can traverse the planet because of their small interaction cross section with matter\footnote{Neutrino absorption in the Earth becomes relevant only above a few tens of TeV.}. However, the pattern of Cherenkov photons detected in the instrumented volume and emitted by downward-going muons could occasionally mimic the signal produced by upward-going neutrino interactions. The first step of any neutrino search thus consists in removing these \textit{misreconstructed} muons. Once these events are removed, most of the neutrino events detected by a neutrino telescope will be due to \textit{atmospheric neutrinos}, also coming from cosmic ray extensive air showers. Only a small contribution in data comes from the cosmic neutrino signal. 

The high-purity neutrino sample that will be used in the following is prepared by selecting events on the basis of different quality criteria defined from the output of the event reconstruction algorithms, separately for each event topology. These criteria are defined on simulated Monte Carlo datasets~\cite{bib:antares_mc} which, in the \textit{run-by-run} approach, are prepared taking into account the environmental conditions present at each moment of data acquisition, including the ageing of the detector~\cite{bib:antaresk40}. The optimisation of selection cuts is done \textit{blindly}, that is without looking at the entirety of data to avoid biases in the selection procedure. In the Monte Carlo datasets, both atmospheric muons passing through the detector and neutrino interactions are simulated. Atmospheric muons are simulated with the \texttt{mupage} code~\cite{bib:mupage}, which uses parametric formulae~\cite{bib:mupage_par} providing an estimation of the muon flux, of the muon multiplicity of atmospheric muon bundles, and of the muon lateral distribution at large depths under the sea. Neutrino interactions are simulated using an \textit{ad hoc} software developed in the ANTARES Collaboration~\cite{bib:antares_mc}. 

\subsubsection{Track events}

The rejection of atmospheric muons is straightforward in the track sample, and is achieved requiring an upward-going reconstructed direction of the event ($\theta_{\textrm{zen}} > 90^\circ$). The track reconstruction algorithm provides two output parameters: the reconstruction quality $\Lambda$, measured as the maximum of the likelihood function from the reconstructed algorithm over the number of degrees of freedom of the fit, and the estimated angular error in the direction reconstruction, $\beta$. Misreconstructed atmospheric muons are strongly suppressed by requiring appropriate values of $\Lambda$ and a value of $\beta<0.5^\circ$ (the reader can refer to~\cite{bib:antares_ps} for additional details). Overall, 3392 neutrino events survive this selection in 4541 days of analysed effective livetime, 99\% of which are expected to be originating from cosmic ray interactions in the atmosphere, and about 1\% of which should be of cosmic origin, according to the simulations for a reference power-law spectrum (as defined in equation~\ref{eq:powerlaw}) with $\gamma = 2.4$ and $\phi_{\textrm{astro}}=1.0$. The contamination of atmospheric muons in this sample, estimated from Monte Carlo simulations, is below 0.3\%. This sample is extremely pure also in its neutrino flavour composition, since 99.9\% of the selected events are muons originating from $\nu_\mu$ charged current interactions. The simulated neutrino rate in Monte Carlo simulations is 25\% smaller than the detected neutrino rate below 1~TeV, when considering the reference flux from~\cite{bib:honda}; this is in agreement with previous ANTARES results~\cite{bib:antares_atmo_tr}, and is within the systematic uncertainties in atmospheric neutrino models~\cite{bib:atmo_syst}. More recent models of the atmospheric neutrino flux are available~\cite{bib:honda_new}; however, including them would not change the outcome of this analysis since the event sample collected with ANTARES does not allow for a statistical discrimination between the spectral features of each model. The overall atmospheric neutrino flux normalisation will be one of the sources of systematic uncertainties that will be considered in the statistical analysis.%Even though more modern models of the atmospheric neutrino flux are available, the event sample collected with ANTARES is not sensitive enough to be affected by possible differences; the atmospheric neutrino flux will be one of the sources of systematic uncertainties that will be considered in the statistical analysis.

\subsubsection{Shower events}

The selection of showers is more difficult than the track selection, since downward-going atmospheric muons can undergo catastrophic energy losses producing electromagnetic showers, which will appear in the detector very similar to neutrino-induced showers. If the interacting neutrino has an energy above $\sim1$~TeV, a pure neutrino sample can be obtained with a series of selection cuts in a similar way as for the track sample. First of all, events passing the selection criteria for tracks are excluded, so that independent samples can be built for a combined statistical analysis (see section~\ref{sec:stat}). Subsequently, a set of minimal criteria must be satisfied: the reconstructed shower direction is below the horizon; the reconstructed interaction vertex is within a fiducial cylinder, 300~m in radius and 500~m in height, centred at the centre of gravity of the detector; the goodness of fit of the vertex position estimation is good (M-estimator value~$< 1000$); a likelihood-ratio test between the shower and the track hypotheses for the measured signals on the photomultipliers favours the shower hypothesis~\cite{bib:antares_tantra}. Then, the optimal cut is chosen by requiring that the angular error estimation from the directional reconstruction is better than $10^\circ$, that the likelihood-ratio test mentioned above strongly favours the shower hypothesis, and that a Random Decision Forest algorithm~\cite{bib:antares_dusj} also strongly favours the signal hypothesis (neutrino-induced shower event) against the background (atmospheric muon event). After this selection, 187 events are observed in 4541 days of analysed livetime, more than 95\% of which are neutrino-induced events, according to Monte Carlo simulations. These neutrino events are split into the different flavours, and about 8\% of them should be of cosmic origin, as from simulations with a power-law spectrum with $\gamma = 2.4$ and $\phi_{\textrm{astro}}=1.0$, with electron neutrinos being twice as abundant as the other neutrino flavours. Differently from the track sample, Monte Carlo estimations for the atmospheric neutrino component are above the observations in data by $\sim10$\%, within systematic uncertainties and as already observed in previous analyses~\cite{bib:antares_atmo}.

\subsubsection{Low-energy shower events}

The selection of events in the TeV energy range and below for the shower topology becomes even more difficult. A sample of such low-energy showers can be however very important to complement the information coming from higher-energy events. For this reason, a dedicated Boosted Decision Tree classifier has been developed~\cite{bib:antares_atmo} to select TeV neutrinos against atmospheric muons. At first, events passing the selection for the track and shower samples defined above are excluded, so that this third sample is independent from the other two. Then, a similar procedure as for the high-energy showers is applied, requiring a containment in the same fiducial volume, and a good reconstruction quality, using the same variables described above. Finally, a selection cut on the Boosted Decision Tree classifier score is applied to produce a high-purity neutrino sample: 219 events pass this selection in data in the analysed 4541 days of livetime. At least 99\% of these events are neutrino-induced, according to simulations, and about 2.5\% of them should be of cosmic origin for the reference power-law spectrum with $\gamma = 2.4$ and $\phi_{\textrm{astro}}=1.0$; also in this case, electron neutrinos are more abundant than other flavours. Monte Carlo estimations are below data observations by about 10\% for this sample, within the assumed systematic uncertainties. 

\begin{figure}
  \centering
  \begin{minipage}{0.6\textwidth}
    \includegraphics[width=\textwidth]{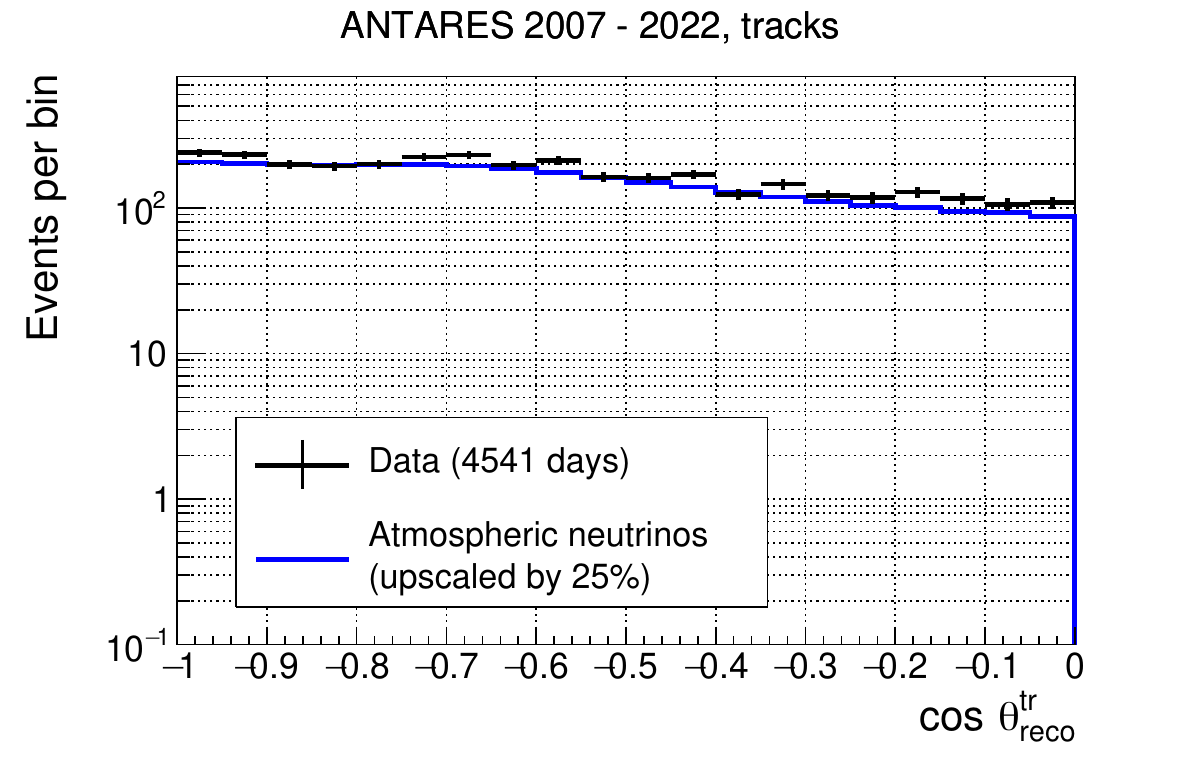}
  \end{minipage}

  \begin{minipage}{0.6\textwidth}
    \includegraphics[width=\textwidth]{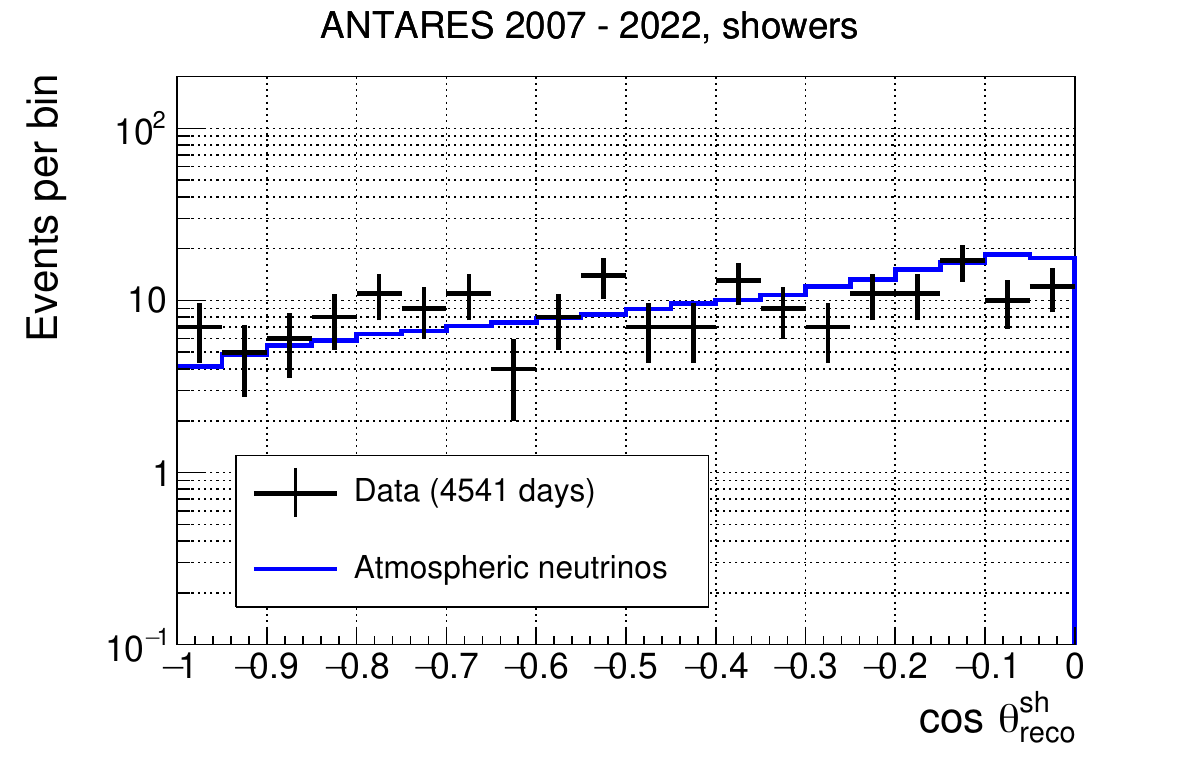}
  \end{minipage}

  \begin{minipage}{0.6\textwidth}
    \includegraphics[width=\textwidth]{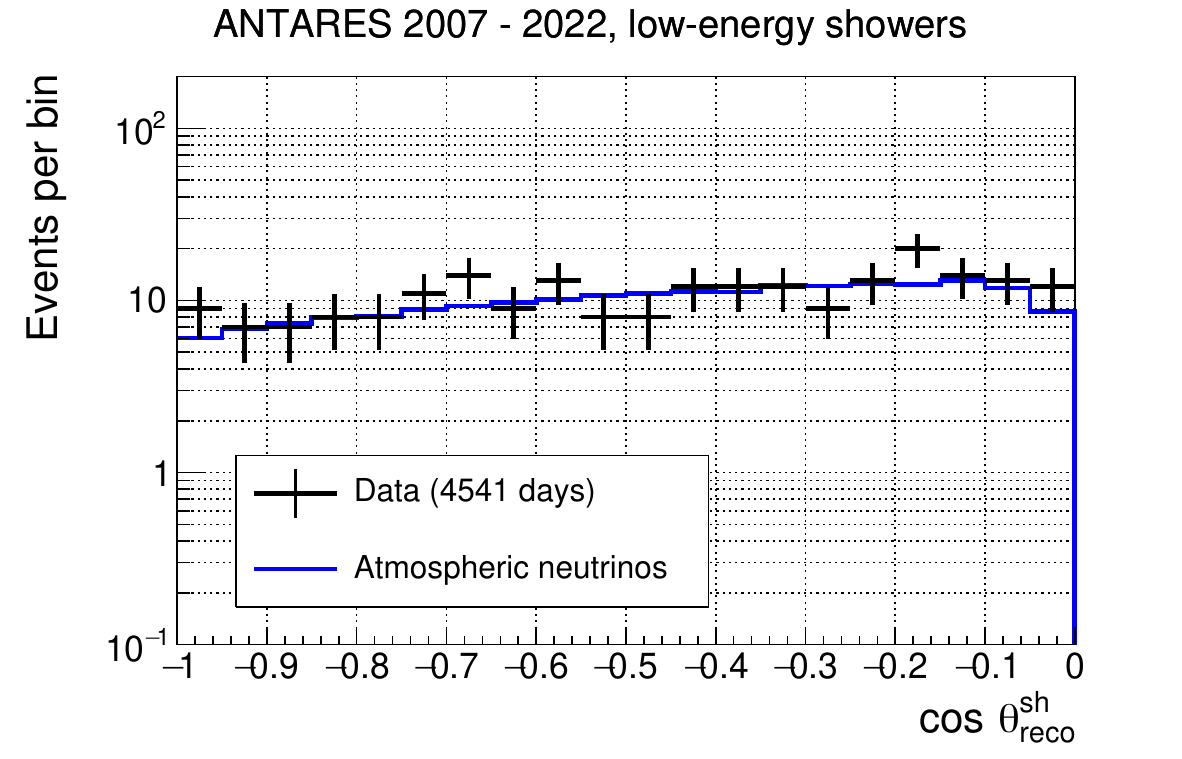}
  \end{minipage}
  \caption{Distribution of the cosine of the reconstructed zenith angle $\theta_{\textrm{reco}}$ for Monte Carlo simulations of atmospheric neutrinos (blue lines), compared with data (black crosses). Events selected as tracks are shown in the top row, events selected as showers are in the middle row, and events selected as low-energy showers in the bottom row. Monte Carlo simulations are upscaled by 25\% in the track sample, while the nominal normalisation is assumed for the two shower samples.}
  \label{fig:data_mc}
\end{figure}

A summary of the comparison between data and Monte Carlo simulations for atmospheric neutrinos is provided in figure~\ref{fig:data_mc} for the reconstructed zenith angle of the selected events. The simulation of atmospheric neutrinos is upscaled by 25\% for tracks, while no scaling is applied for showers.

\section{Statistical analysis}\label{sec:stat}

A diffuse cosmic component can be identified against atmospheric events on the basis of the distribution of the estimated energy values $E_{\textrm{reco}}^{tr/sh}$, with $tr$ indicating the track reconstruction and $sh$ indicating the shower reconstruction algorithm. The energy spectrum of cosmic neutrinos is harder than that of atmospheric neutrinos, whose spectral index is $\gamma_{\textrm{atm}}\simeq 3.6$; because of this, the cosmic component will yield more events in the high-energy tail of the estimated energy distributions. A good energy resolution is fundamental in differentiating between these components. Even though the energy reconstruction is optimal for showers, events classified as tracks are in any case useful because of the larger detection efficiency in this sample, coming from the larger volume inside which they can be detected and selected.

Following the same statistical approach used in the search for neutrinos from the Galactic Plane~\cite{bib:antares_gal}, the estimated energy distributions are binned and analysed to extract the parameters (as in equation~\ref{eq:powerlaw}) that best describe the properties of the cosmic neutrino flux. The following likelihood function is defined
\begin{equation}
  \mathcal{L}\left(N_i, S_i^{(\gamma)}, B_i, \phi_{\textrm{astro}} \right) = \prod_{k}\prod_{i=1}^{ N^{k}_{\textrm{bins}}}\mathcal{P}(N_i, B_i + \phi_{\textrm{astro}}S_i^{(\gamma)}).
  \label{eq:fitting}
\end{equation}
In this equation, the term $\mathcal{P}(N_i, B_i + \phi_{\textrm{astro}}S_i^{(\gamma)})$ represents the Poisson probability of having $N_i$ events in the $i$-th bin of the data distribution, with $B_i$ being the corresponding expected background in Monte Carlo simulations for that bin, and $S_i^{(\gamma)}$ being the signal prediction for a given spectral index $\gamma$ in the same bin. The number of signal events in each bin is scaled by the normalisation factor $\phi_{\textrm{astro}}$. The product runs over the number of bins of the energy distribution for the three samples $k$, namely tracks, showers, and low-energy showers. The spectral index $\gamma$ can take values within $[1.5, 3.5]$ with steps of 0.05; the signal normalisation is varied as $\phi_{\textrm{astro}}\in [10^{-2}, 10]$ with $\log_{10}\phi_{\textrm{astro}}$ steps of 0.02.

A Bayesian statistical treatment is applied to compute the posterior probability in the $(\phi_{\textrm{astro}}, \gamma)$ phase space. Statistical and systematic uncertainties on the background and on the signal estimates are included as Gaussian priors $\pi(B_i)$ and $\pi(S_i)$. In addition to the theoretical uncertainty on the atmospheric neutrino flux, additional sources of systematics arise from the optical properties of water and from the overall detection efficiency of the optical modules. These uncertainties have been estimated using Monte Carlo simulations in which the detection efficiency has been modified within the known constraints on the optical properties of water and the optical module response. As a consequence of these studies, the uncertainty on the normalisation of the atmospheric flux is assumed to be 30\%, while the uncertainty on the signal computed with nominal values of the parameters is estimated to be 20\%. A flat prior is considered for the parameters of interest $\phi_{\textrm{astro}}$ and $\gamma$. Finally, the marginalised posterior distribution $P(\phi_{\textrm{astro}}, \gamma)$ is obtained by factoring in the likelihood and the priors, and then integrating
\begin{eqnarray}
  \nonumber      P(\phi_{\textrm{astro}}, \gamma) &=& \int  \Bigl \{  \mathcal{L}\left(N_i, S_i^{(\gamma)}, B_i, \phi_{\textrm{astro}} \right)  \\ \nonumber &\times& \pi(B_i)\times \pi(S_i) \times \pi(\phi_{\textrm{astro}}, \gamma) \\&\times& \prod\left(dB_i dS_i^{(\gamma)}\right) \Bigr\}.
  \label{eq:posterior}
\end{eqnarray}

The procedure is tested using fake datasets produced from Monte Carlo simulations for various values of $\phi_{\textrm{astro}}$ and $\gamma$. In these fake datasets, statistical fluctuations are introduced independently for each sample. The energy range and the binning of the histograms for which the likelihood function of equation~\ref{eq:fitting} is computed are optimised in this testing procedure. The histograms cover the range 300~GeV~--~3~PeV in the estimated energy for all samples; this energy range is binned in equally-spaced bins in the $\log_{10}$ of the estimated energy values; 16 bins are used for the track channel, while 12 bins are used for each of the two shower channels. %A larger number of bins is chosen for tracks because of the larger acceptance for this channel.

\begin{figure}
  \centering
  \begin{minipage}{0.7\textwidth}
    \includegraphics[width=\textwidth]{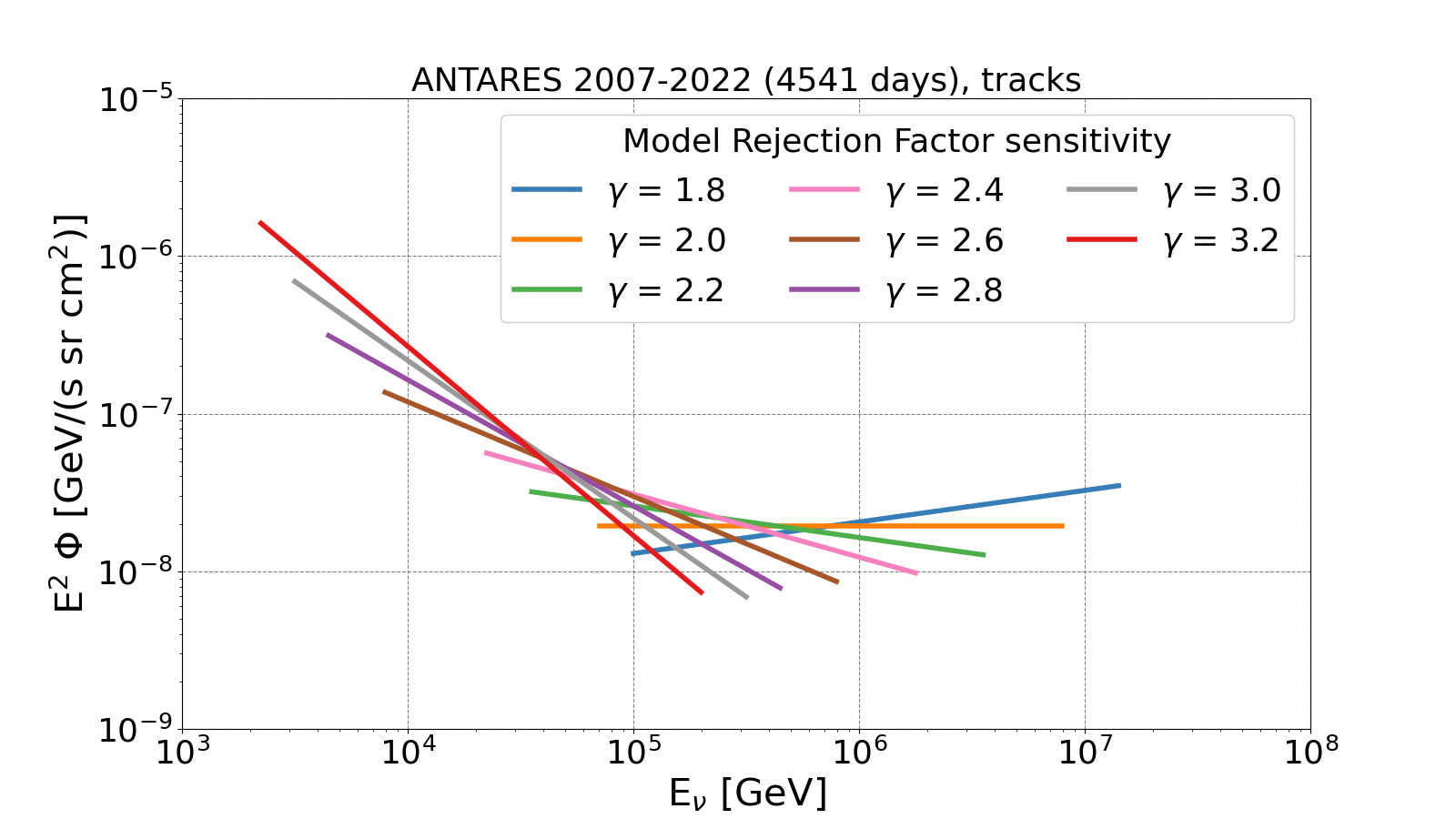}
  \end{minipage}
  
  \begin{minipage}{0.7\textwidth}
    \includegraphics[width=\textwidth]{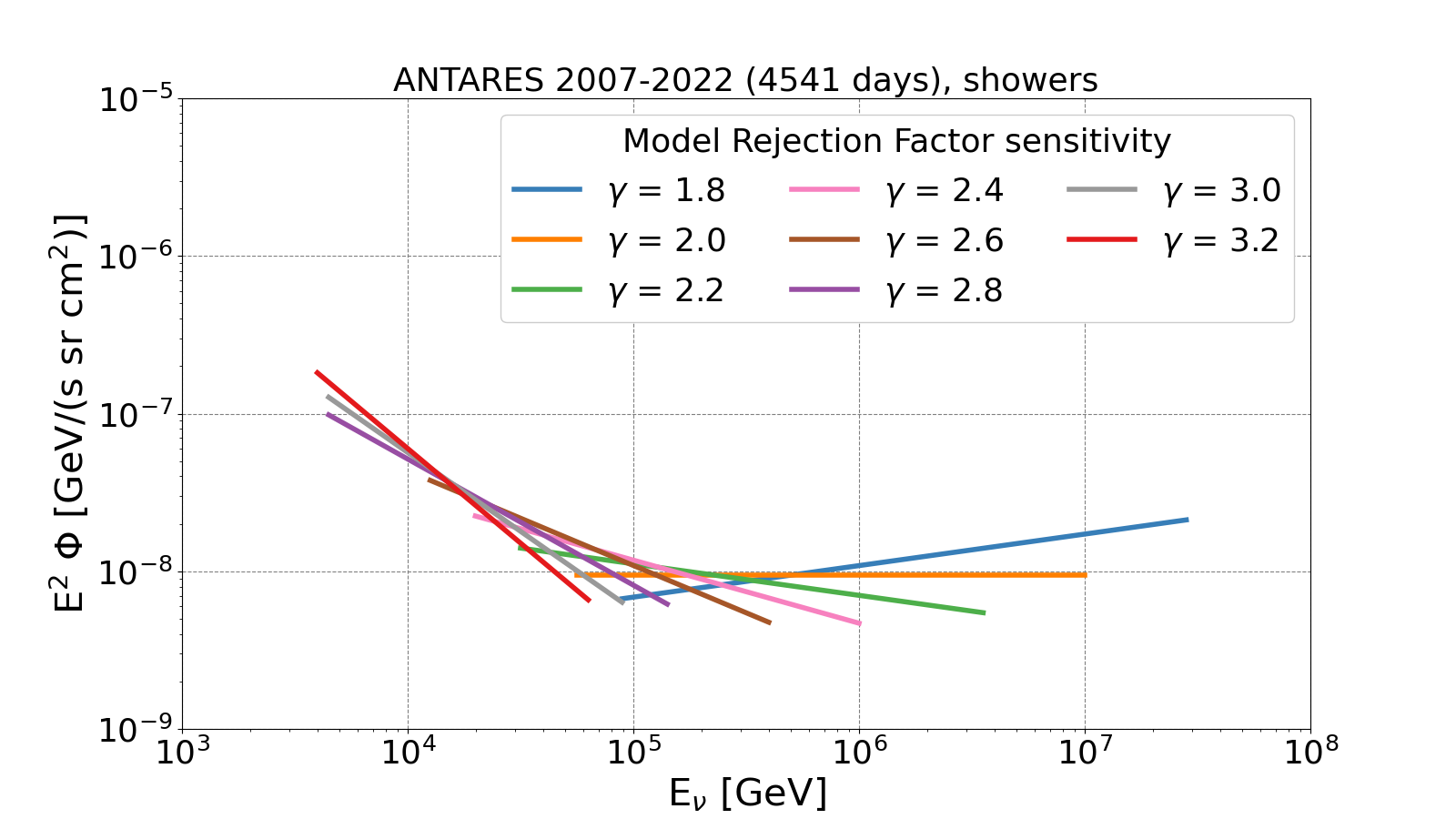}
  \end{minipage}

  \begin{minipage}{0.7\textwidth}
    \includegraphics[width=\textwidth]{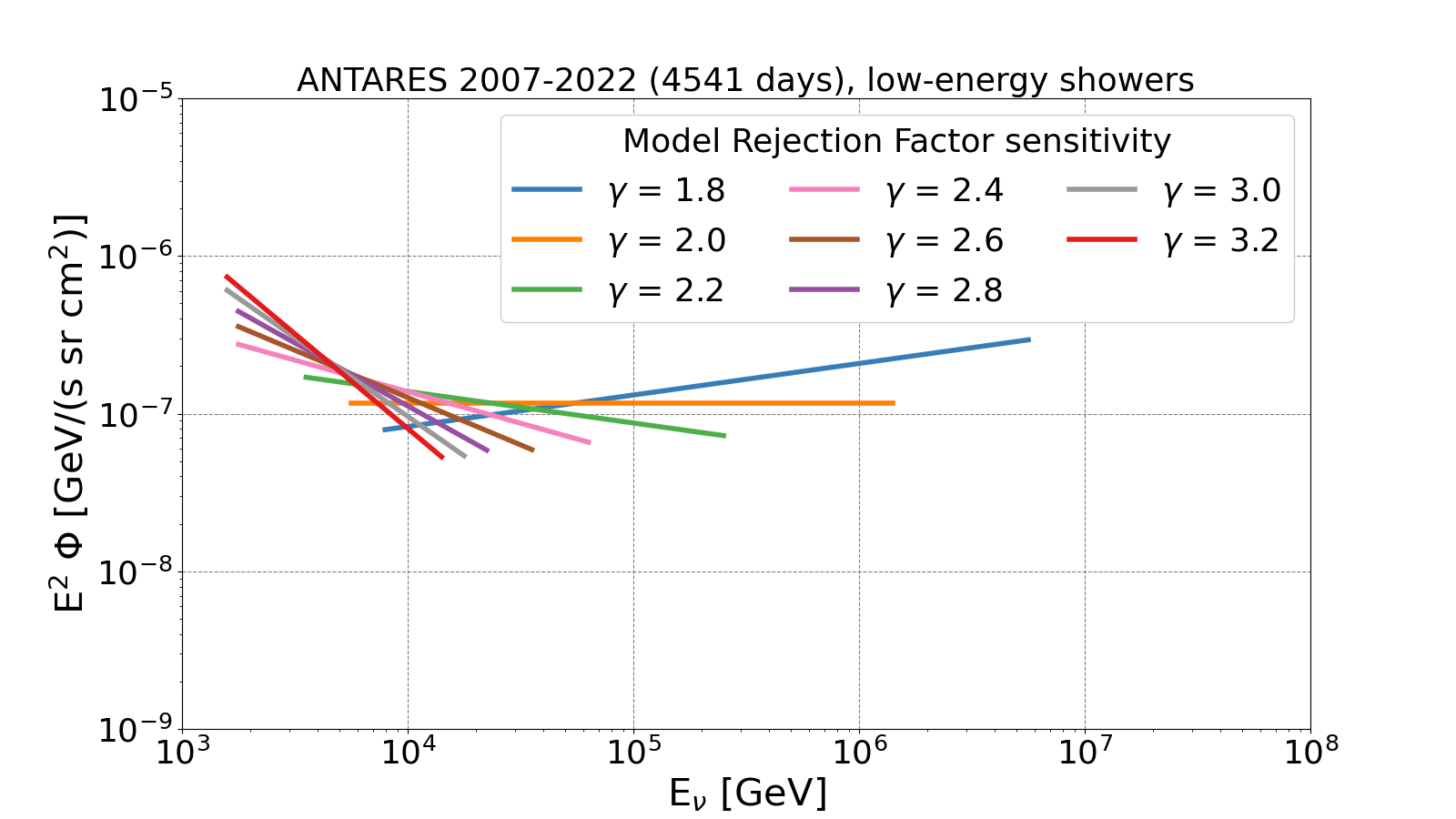}
  \end{minipage}
  \caption{Sensitivity of the analysis samples to a diffuse flux of cosmic neutrinos described by a power-law spectrum as in equation~\ref{eq:powerlaw}, for different spectral indexes $\gamma$ shown in the legend. The three samples are shown separately, from top to bottom: tracks, showers, and low-energy showers. For each spectral index, the lower and upper limit of the energy range corresponds to the 5\% and 95\% quantiles of the corresponding neutrino energy distributions in Monte Carlo simulations.}
  \label{fig:sensy}
\end{figure}

For each value of the tested spectral indexes, an assessment of the sensitivity of each sample is also carried out using the Model Rejection Factor procedure~\cite{bib:mrf},  based on the Feldman and Cousins upper limit estimation in counting experiments~\cite{bib:fc}. In this procedure, the average 90\% confidence level upper limit obtained in the background-only case is defined as the sensitivity of the experiment. The reconstructed energy range which provides the best sensitivity is then estimated for each spectral index and each sample, individually. The 5\% and 95\% quantiles of the corresponding neutrino energy distributions in Monte Carlo simulations are taken as the lower and upper limit of the energy range of validity for the obtained results. These sensitivities are shown in figure~\ref{fig:sensy} for the three samples separately and for selected values of the spectral index. The shower sample is the most sensitivive to the cosmic neutrino signal, reaching normalisation values for the cosmic signal $\phi_{\textrm{astro}}$ as low as 1.0 for most spectral indexes. The sensitivity of the other samples is a factor of 2 to 3 worse but they allow to complement the outcome from the shower sample, in particular below 10~TeV and for soft spectral indexes. The low-energy shower sample becomes particularly relevant in the TeV range for spectral indexes $\gamma > 2.8$, where its sensitivity matches that of the standard shower sample. The energy range of validity for the combination of the three samples is finally obtained by merging the Monte Carlo distribution for the true neutrino energy into a single distribution, for each spectral index, and considering the 5\% and 95\% quantiles. 

The discovery potential of these samples has also been estimated using fake datasets. A cosmic neutrino flux described by the power-law spectrum reported by the IceCube Collaboration for the track channel~\cite{bib:ic_tracks} would yield an observation at $2.5\sigma$ significance level in the combined samples. For a softer spectrum ($\gamma = 2.8, \phi_{\textrm{astro}} = 2$), a $3\sigma$ significance could be reached for the unbroken power-law hypothesis.

\section{Results}\label{sec:results} 

\begin{figure}
  \begin{center}
    \begin{minipage}{0.7\textwidth}
      \includegraphics[width=\textwidth]{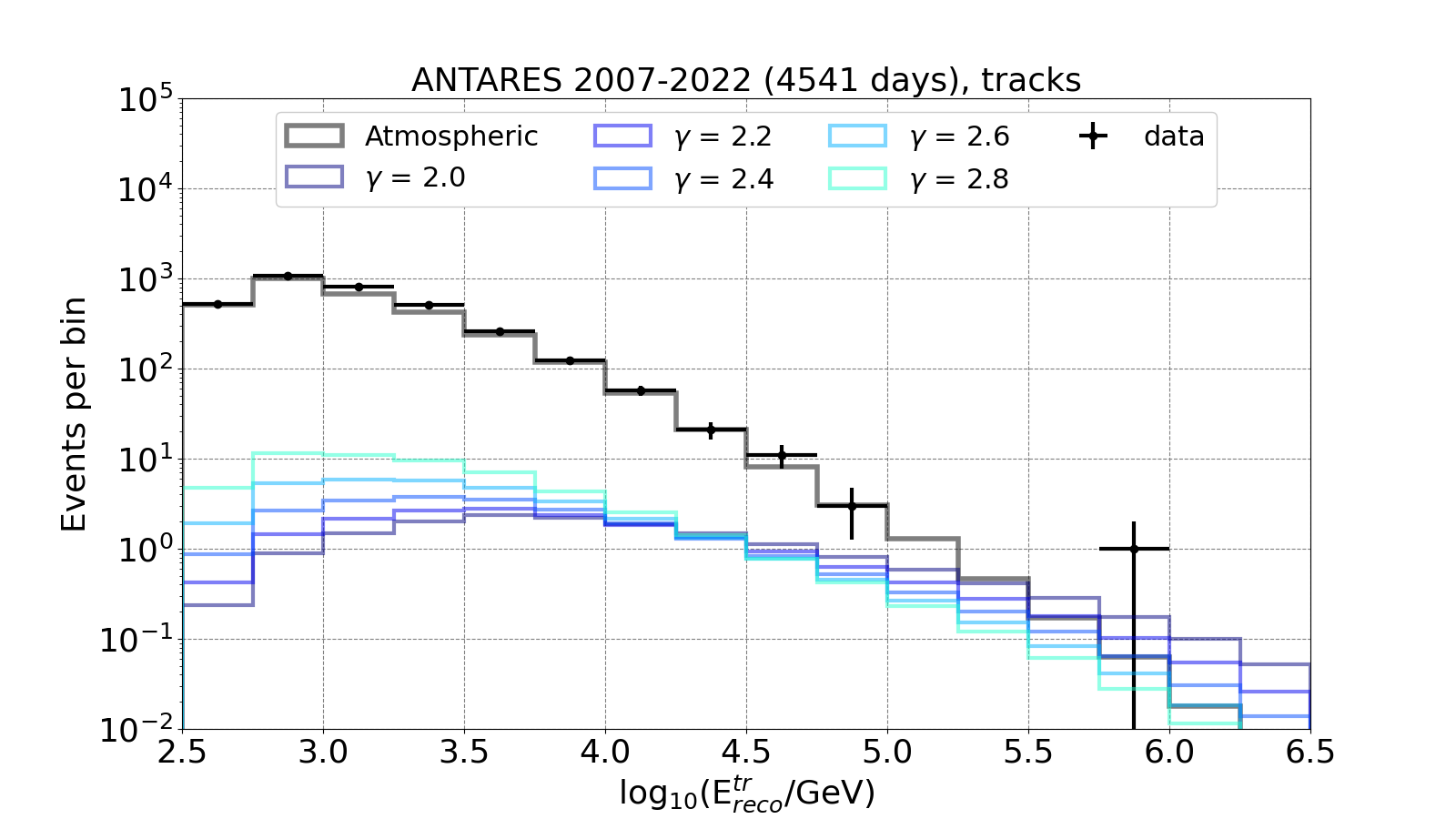}
    \end{minipage}

    \begin{minipage}{0.7\textwidth}
      \includegraphics[width=\textwidth]{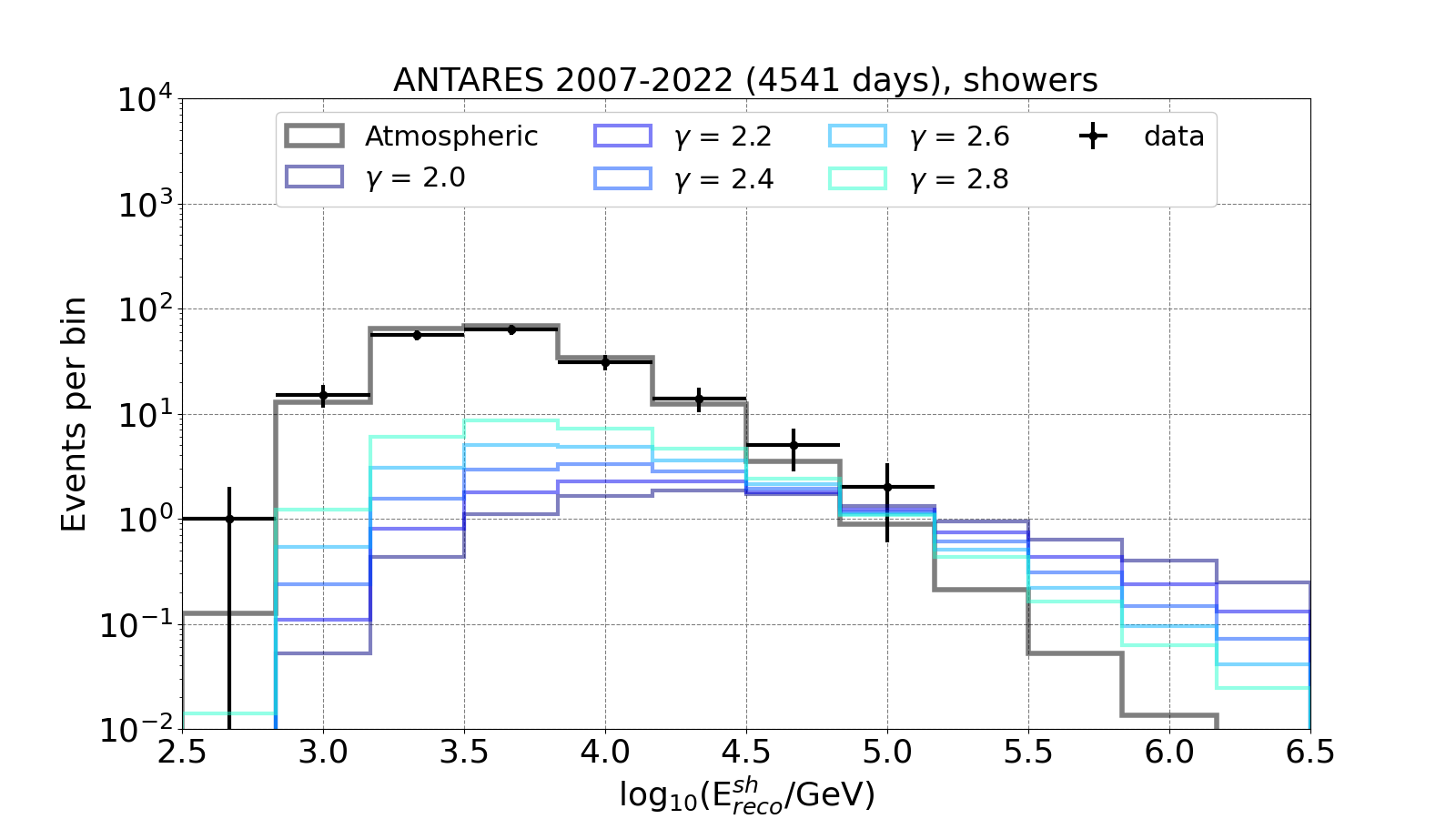}
    \end{minipage}

    \begin{minipage}{0.7\textwidth}
      \includegraphics[width=\textwidth]{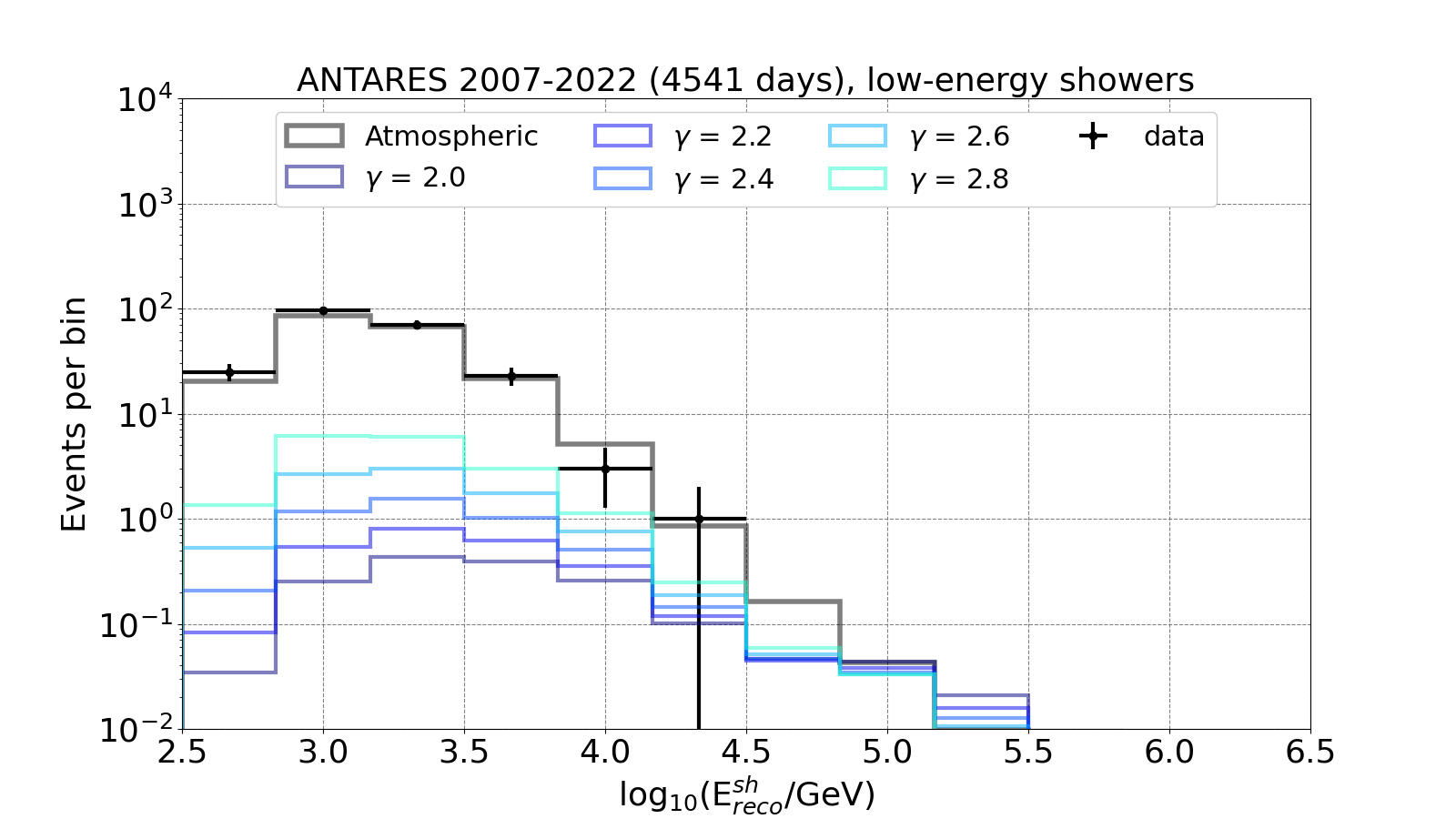}
    \end{minipage}
  \end{center}
  \caption{Distribution of the reconstructed energy for the track (top), shower (middle), and low-energy shower (bottom) sample. Data are represented by black crosses, with statistical errors. The expected atmospheric neutrino distribution from Monte Carlo simulations is shown in grey ---  in the track channel this distribution is upscaled by 25\%; no scaling is applied for showers. The expected signal distributions for a cosmic normalisation $\phi_{\textrm{astro}}=1.0$ and different spectral indexes $\gamma$ (as in equation~\ref{eq:powerlaw}) are shown with the different shades of blue reported in the legend.}
  \label{fig:unblinded_ereco}
\end{figure}

The distribution of the reconstructed energy $E_{\textrm{reco}}^{tr/sh}$ for experimental data are shown in figure \ref{fig:unblinded_ereco} for tracks (top), showers (middle) and low-energy showers (bottom), and compared to simulations. The binning and energy range used in these plots are the same as those fixed for the parameter estimation procedure. In particular, considering the energy cut that from the Model Rejection Factor procedure provides the best sensitivity to a cosmic flux with spectral index $\gamma = 2.4$:
\begin{itemize}
\item Selecting track events with a value of the energy estimator $E_{\textrm{reco}}^{tr}$ larger than 30~TeV, 17 events are observed in data, while 14.3 are expected from the simulations of atmospheric neutrinos (after scaling up the atmospheric neutrino distribution by 25\%). 
\item Selecting showers which have an energy estimator $E_{\textrm{reco}}^{sh}$ larger than 20~TeV, 13 events are observed in data, while 10.3 are expected from the simulations of atmospheric neutrinos.
\item Finally, for the low-energy shower sample, no significant excess is observed, either. When selecting events with energy estimator $E_{\textrm{reco}}^{sh}$ above 2~TeV, 73 events are observed against an expected 60.8 events from the simulations of atmospheric neutrinos. This data-Monte Carlo discrepancy mainly appears at low energies, namely in the bin around 1~TeV, while at higher energies data and Monte Carlo simulations for atmospheric neutrinos match very well.
\end{itemize}

% \subsubsection*{Comparison with previous ANTARES results}

In previous analyses of ANTARES data in the search for a diffuse flux of cosmic neutrinos a $1.8\sigma$ excess of high-energy events was observed~\cite{bib:antares_diffuse, bib:antares_diffuse_icrc}. The low significance of the previous observation is compatible with the present result, for which the significance of the high-energy excess is smaller than $1\sigma$. It is in any case worth mentioning what causes this difference:
\begin{itemize}
\item\textit{Tracks}. Exactly the same event selection cuts are applied in this analysis, with only the addition of about 4 years of data with respect to the latest published results~\cite{bib:antares_diffuse_icrc}. However, an improved energy estimator is currently used: the energy estimation for tracks has been indeed calibrated with the newest and most accurate Monte Carlo simulations, to account for the time-dependent efficiency loss of the detector. %This may have cured possible systematic effects in the energy estimation that were present previously, which may have lead to some fluctuations in the tail of the distribution.
\item\textit{Showers}. The excess of high-energy events in the shower sample has also decreased. In this case, a new event selection procedure with more stringent cuts has been applied, leading to a purer neutrino sample. The reduction by at least a factor of 5 in the number of atmospheric muons contaminating the neutrino sample might have caused the decrease in the number of events in the high-energy tail. This new selection was done blindly, as already mentioned, without looking at data and was driven by the necessity of reducing the systematic uncertainties connected with the estimation of the rate of surviving atmospheric muons. %The outcome of the new analysis is thus consistent with the new selection choice.
\end{itemize}

\subsection{Constraints on the single power-law assumption} \label{sec:fit_res}

\begin{figure}
  \centering
  \includegraphics[width=\textwidth]{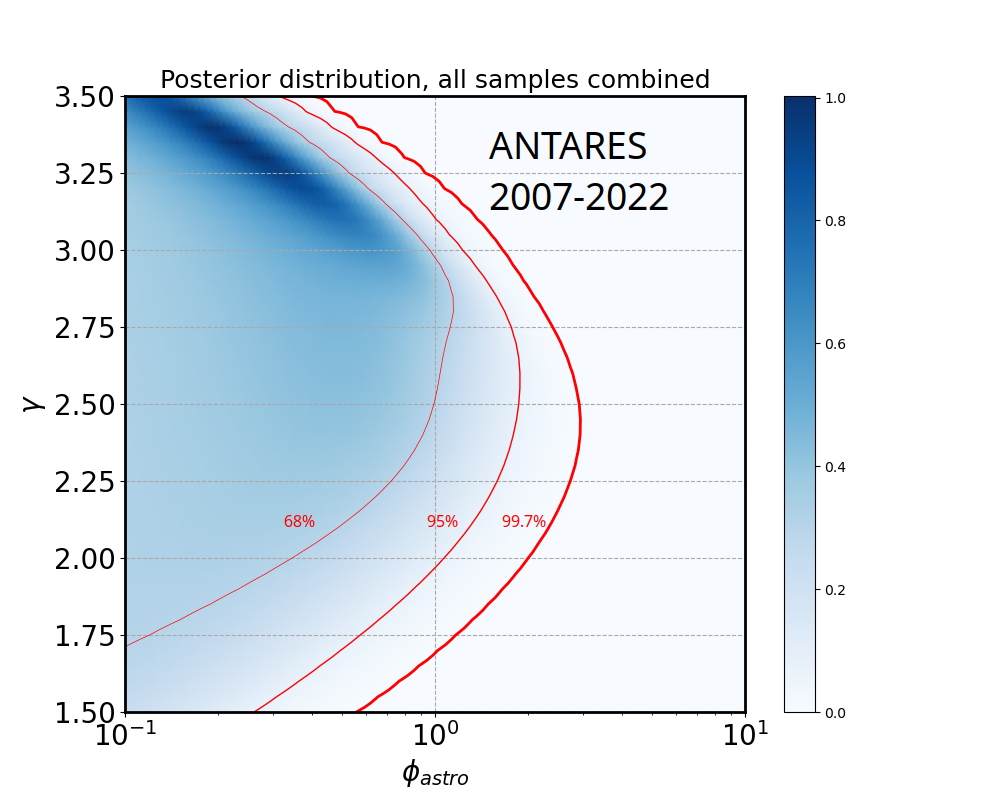}
  \caption{Posterior probability distribution (see equation~\ref{eq:posterior}) in the 2D ($\phi_{\textrm{astro}}, \gamma$) phase space from the fit of all ANTARES samples combined together. The colour scale shows the posterior probability upscaled so that the maximum is set to 1. The three red lines show (from thinnest to thickest) the 68\%, 95\%, and 99.7\% posterior probability credible area.}
  \label{fig:2d_fit_comb}
\end{figure}

The reconstructed energy distributions from the three samples are fitted together using the procedure described in the previous section to determine the posterior probability in each point of the analysed phase space $(\phi_{\textrm{astro}}, \gamma)$; the point in the phase space with the highest value of the posterior probability can be considered the best-fit point. The posterior distribution is reported in figure~\ref{fig:2d_fit_comb}, with a maximum at (0.23, 3.35): the low flux normalisation and the fact that the best-fit spectral index is close to that of atmospheric neutrinos indicate that the observed reconstructed energy distribution is compatible with the atmospheric neutrino flux. The computed p-value of the excess at the best-fit point is $0.3$, corresponding to a $0.55\sigma$ significance (one-sided convention).

\begin{figure}
  \centering
  \includegraphics[width=\textwidth]{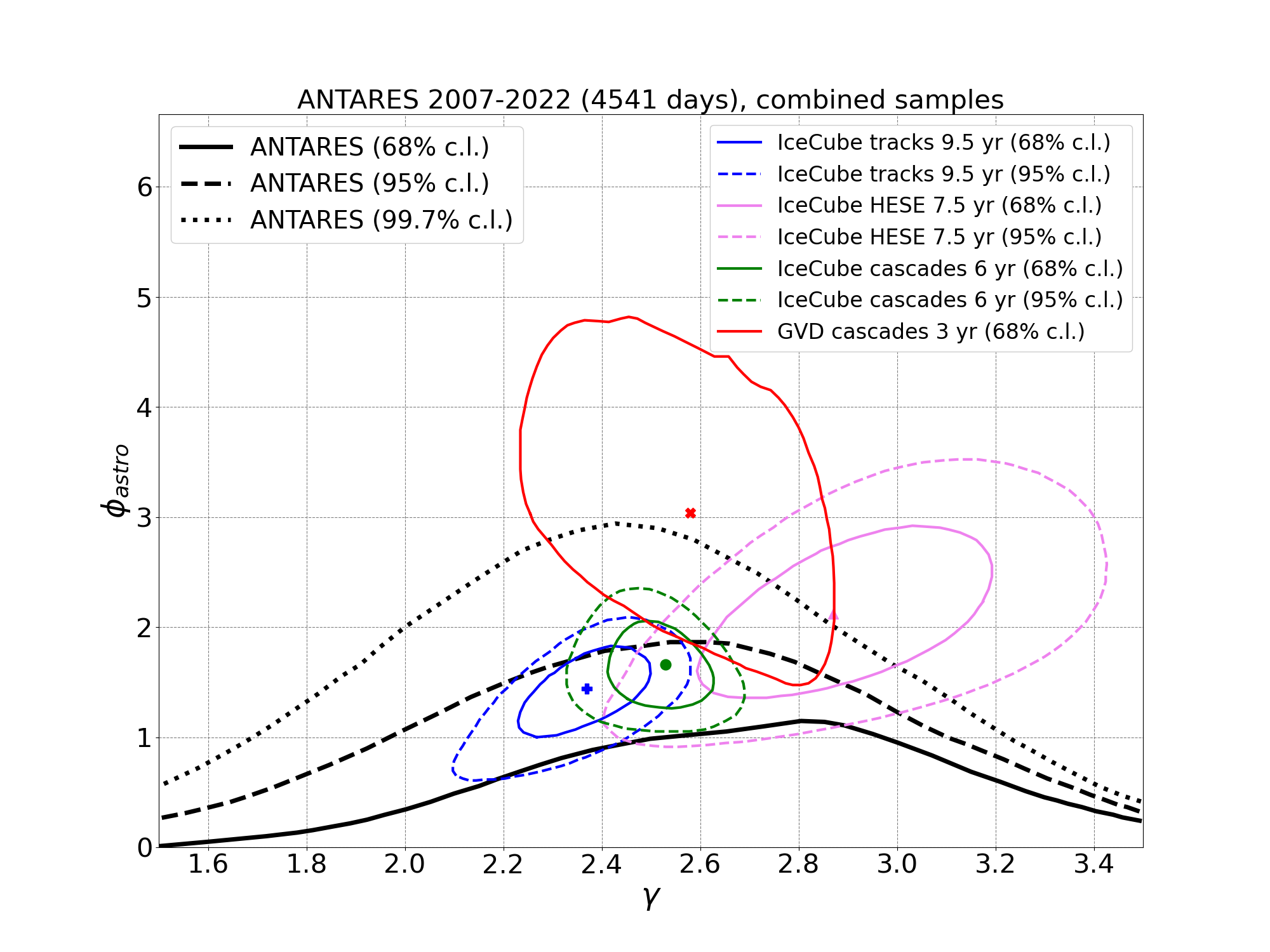}
  \caption{Contours at 68\% (solid) and 95\% (dashed) confidence level from IceCube analyses (HESE~\cite{bib:ic_hese_latest} in pink, tracks~\cite{bib:ic_tracks} in blue, cascades~\cite{bib:ic_cascades} in green) compared to the 68\% (solid), 95\% (dashed), and 99.7\% (dotted) posterior probability credible areas obtained in the combined analysis of the three ANTARES samples (black lines). The IceCube best-fit points are shown with symbols. The Baikal-GVD 68\% confidence level contour and  best-fit point~\cite{bib:gvd_diffuse} are also shown, in red.}
  \label{fig:contours_compared}
\end{figure}

From the posterior distribution, the regions containing 68\%, 95\%, and 99.7\% of the posterior probability can be built, shown as red contours in figure~\ref{fig:2d_fit_comb}. These \textit{credible areas} indicate that there is a 68\%, 95\%, and 99.7\% probability, respectively, that the true physical value is inside such contours. The credible areas obtained here are compared to the 68\% and 95\% confidence level contours from the measurement of the diffuse cosmic flux in IceCube and Baikal-GVD data in figure~\ref{fig:contours_compared}. The $(\phi_{\textrm{astro}}, \gamma)$ best fit from the IceCube track and cascade samples are inside the 95\% credible area, and so are the corresponding 68\% confidence level contours. This agreement is not observed for the IceCube HESE and Baikal-GVD results: both best-fit points are outside of our 99.7\% credible area, and most of the 68\% contours are outside of the 95\% credible area obtained in this work. 

Plots like the one in figure~\ref{fig:contours_compared} only convey a limited amount of information: each analysis reported there is most sensitive in a well defined energy range; each sample is dominated by events arising from different regions of the sky; the various neutrino flavours and interaction channels contribute differently in each of them. Given the limited statistics of the ANTARES sample, the dependency of the posterior probability on the arrival direction of the selected neutrinos and the details of the flavour composition cannot be tested: instead, a check on the energy range of applicability of the fit of an unbroken single power law is straightforward. Fixing a given value of the spectral index $\gamma$, the posterior probability can be profiled to obtain upper limits on the normalisation of the cosmic flux. Table \ref{tab:ul} collects such limits at 68\%, 95\%, and 99.7\% probability for selected values of $\gamma$, with the energy range of applicability of the limits following the definition of the energy range of validity from section~\ref{sec:stat}.

\begin{table}
  \begin{center}
    \begin{tabular}{c | c c  c | l}
      $\gamma$ & $\phi_{\textrm{astro}}^{68\%}$ & $\phi_{\textrm{astro}}^{95\%}$ & $\phi_{\textrm{astro}}^{99.7\%}$  & Energy range \\ & &  & &[TeV]  \\
      \hline
      3.2 & 0.51 & 0.68 & 0.94 & 1.8 -- 63\\
      3.0 & 0.82 & 1.03 & 1.49 & 2.0 -- 100 \\
      2.8 & 0.98 & 1.49 & 2.06 & 2.2 -- 180\\
      2.6 & 0.98 & 1.80 & 2.61 & 2.5 -- 450\\
      2.4 & 0.94 & 1.80 & 2.86 & 2.8 -- 1000\\
      2.2 & 0.78 & 1.64 & 2.73 & 8 -- 2800\\
      2.0 & 0.59 & 1.24 & 2.17 & 30 -- 8000\\
      1.8 & 0.37 & 0.82 & 1.49 & 80 -- 20000
    \end{tabular}
    \caption{The 68\%, 95\%, and 99.7\% probability upper limits on the cosmic flux normalisation, $\phi_{\textrm{astro}}^{\textrm{prob}}$, obtained from profiling the posterior probability for different spectral indexes $\gamma$ are reported. The energy range of validity is provided, following the definition of section~\ref{sec:stat}.}
    \label{tab:ul}
  \end{center}
\end{table}

\begin{figure}
  \centering
  \includegraphics[width=\textwidth]{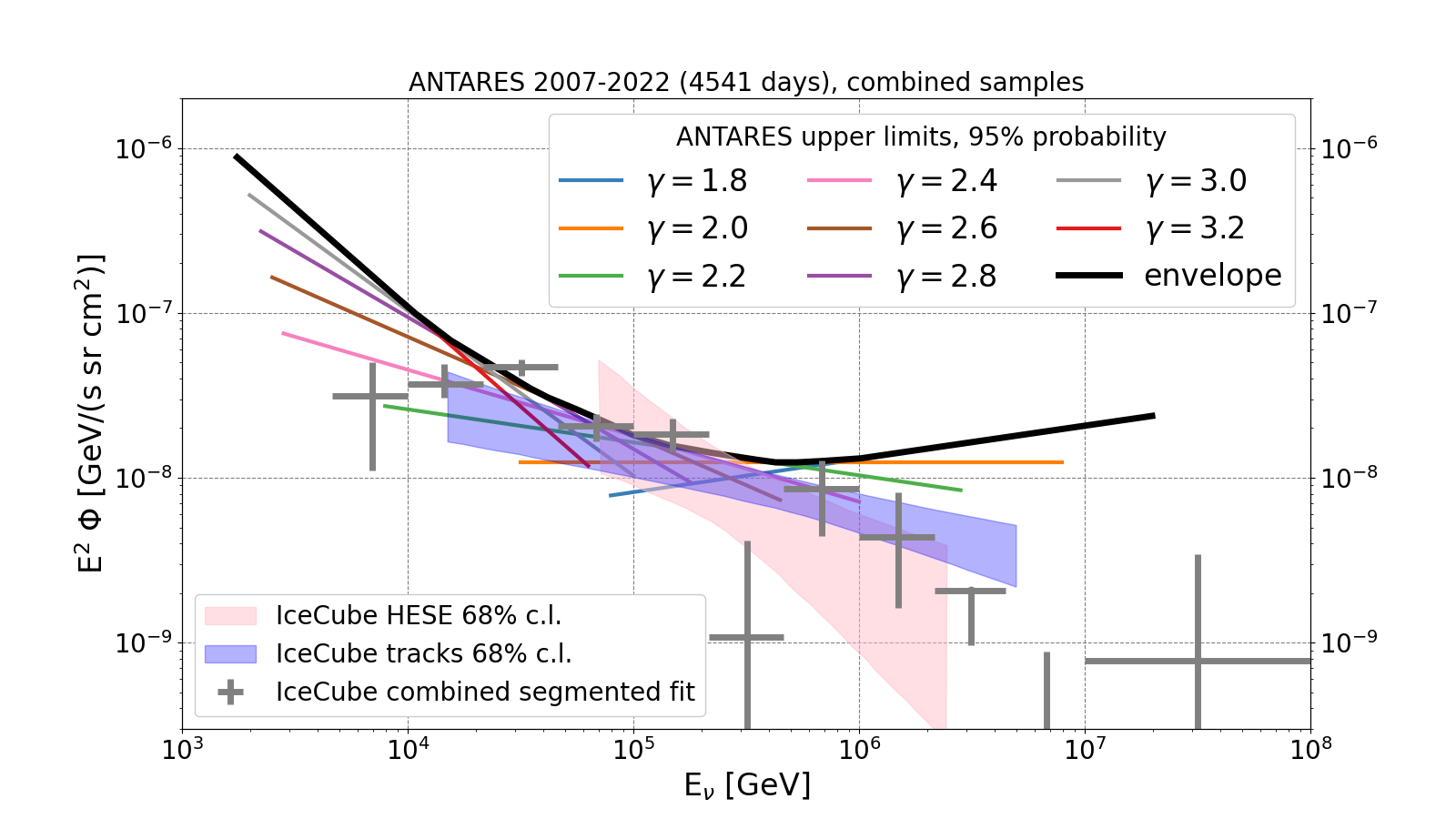}
  \caption{The ANTARES 15-years 95\% probability upper limits for different spectral indexes (coloured lines in the legend) are reported in the figure. The envelope of the limits (black) is taken as the least restrictive limit at every energy. The shaded areas represent the 68\% confidence level intervals for the measurements obtained with the IceCube HESE sample~\cite{bib:ic_hese_latest} in pink and the IceCube track sample~\cite{bib:ic_tracks} in blue. The results from the $E^{-2}$ segmented fit of the IceCube combined samples~\cite{bib:ic_comb} are also shown in grey.}
  \label{fig:segm_lim}
\end{figure}

For soft spectra, the ANTARES results extend to the TeV region, almost one order of magnitude below what has been obtained with IceCube data: this can be attributed to the lower background rates in ANTARES, the larger neutrino detection efficiency in the TeV range, and the specific addition to the analysis of low-energy showers. At these energies, the hypothesis of a single unbroken power-law spectrum may not hold anymore. The combined analysis of different IceCube data samples~\cite{bib:ic_comb}, indeed, shows some preference for a spectral break in the 10~--~30~TeV energy range. The segmented fit of those data gives, for each energy bin in the true neutrino energy, an estimation of the cosmic flux assuming that inside that energy bin the spectrum is compatible with an $E^{-2}$ power law (so that the multiplication by $E^2$ flattens the bin content). A spectral feature is visible around 30~TeV (even though with limited statistical significance). In figure~\ref{fig:segm_lim}, this IceCube result and the 68\% confidence level intervals obtained in the analysis of the IceCube HESE and track samples are compared to the 95\% probability limits reported in table~\ref{tab:ul}. An envelope of the ANTARES limits is also shown, considering for every energy the least restrictive available limit. The tension that could be visible in the comparison of the contours in figure~\ref{fig:contours_compared} is mitigated when accounting for the different energy ranges where each measurement is valid.

\subsection{Study of spectral features} \label{sec:fit_cut}

\begin{figure}
  \centering
  \begin{minipage}{0.9\textwidth}
    \includegraphics[width=\textwidth]{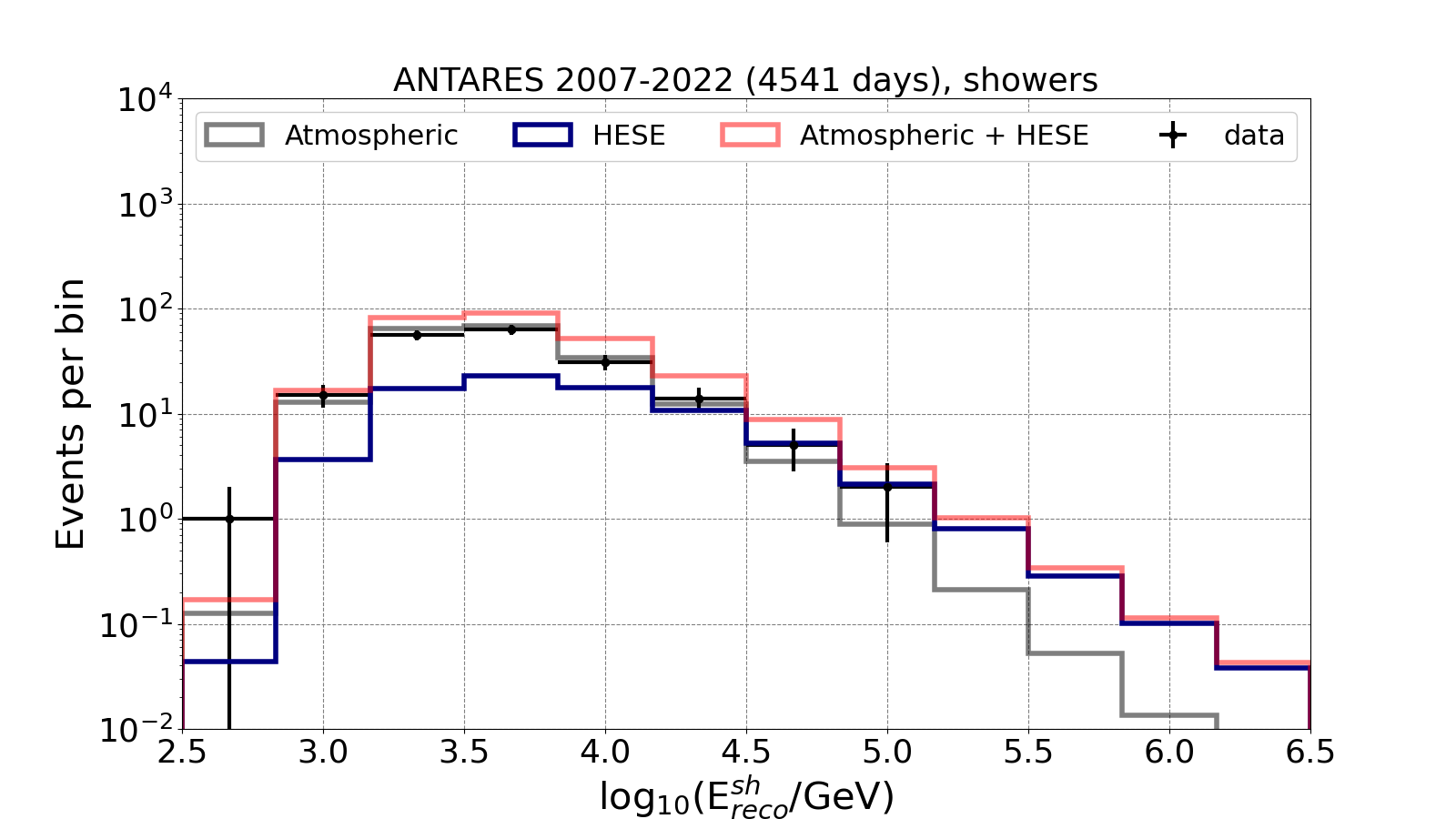}
  \end{minipage}
  \begin{minipage}{0.9\textwidth}
    \includegraphics[width=\textwidth]{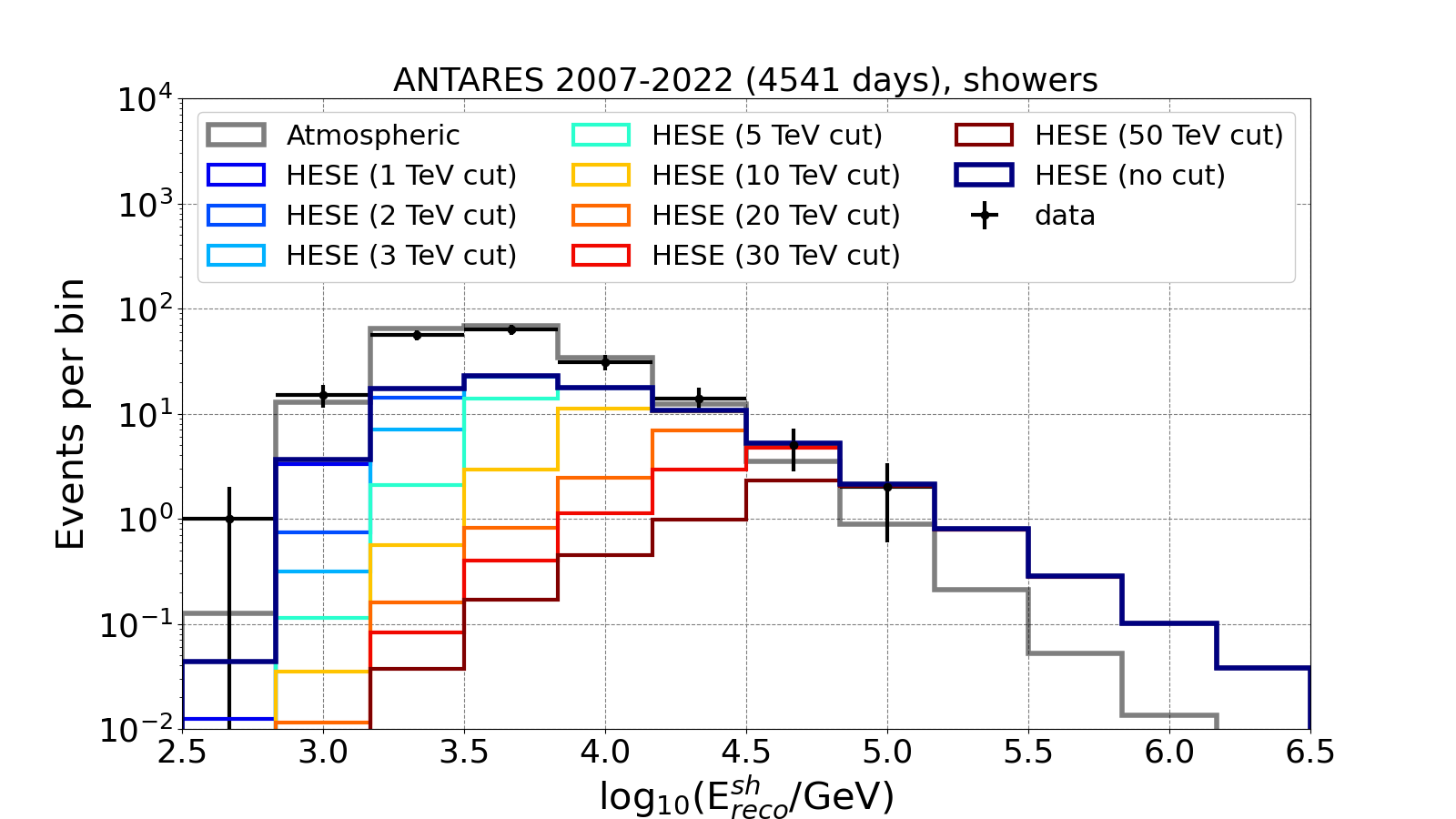}
  \end{minipage}
  \caption{Top: reconstructed energy distribution in the ANTARES shower sample from data (black crosses), compared to the Monte Carlo estimations for atmospheric neutrinos (in grey), and for the IceCube HESE flux (in blue) assuming an unbroken power law; the sum of the expectations from atmospheric neutrinos and cosmic neutrinos following the HESE power-law fit is shown in red. Bottom: data compared to the atmospheric flux expectations and to the HESE power-law fit assuming no cut or a sharp cut-off that removes signal events below different energy thresholds (coloured lines as in the legend).}
  \label{fig:hesediffcut}
\end{figure}

The extension to lower energies of the ANTARES neutrino sample can give useful information on the low-energy features of the cosmic spectrum. This is exemplified in figure \ref{fig:hesediffcut}. In the upper panel of this figure, the expected distribution of the energy estimator for the selected ANTARES showers is shown: data are compared with simulations of the atmospheric neutrino flux, and with a cosmic signal described by an unbroken single power-law spectrum for which the best-fit normalisation and spectral index of the IceCube HESE sample are assumed; in the same plot, also the sum of the two simulated distributions is shown. For values of the shower energy estimator above 100~TeV, the ANTARES data sample has limited power in constraining the signal. However, for energies between 5 and 50~TeV, a clear difference can be observed when comparing the sum of the atmospheric and the HESE component to data. This can qualitatively justify the observed exclusion of the HESE single power-law fit at 99.7\% Bayesian posterior probability. In the bottom panel of the figure, the spectrum is modified assuming a null flux below $E_{\nu}^{\textrm{cut}} = 1, 2, 3, 5, 10, 20, 30, 50$~TeV, and the power-law behaviour as in equation \ref{eq:powerlaw} above $E_{\nu}^{\textrm{cut}}$. The signal distribution is clearly modified if such low-threshold cut-off is present. Only for $E_{\nu}^{\textrm{cut}} > 30$~TeV the sum of the atmospheric plus cosmic signal -- not shown in the figure for simplicity -- becomes qualitatively compatible with data.

\begin{figure}
  \centering
  \includegraphics[width=\textwidth]{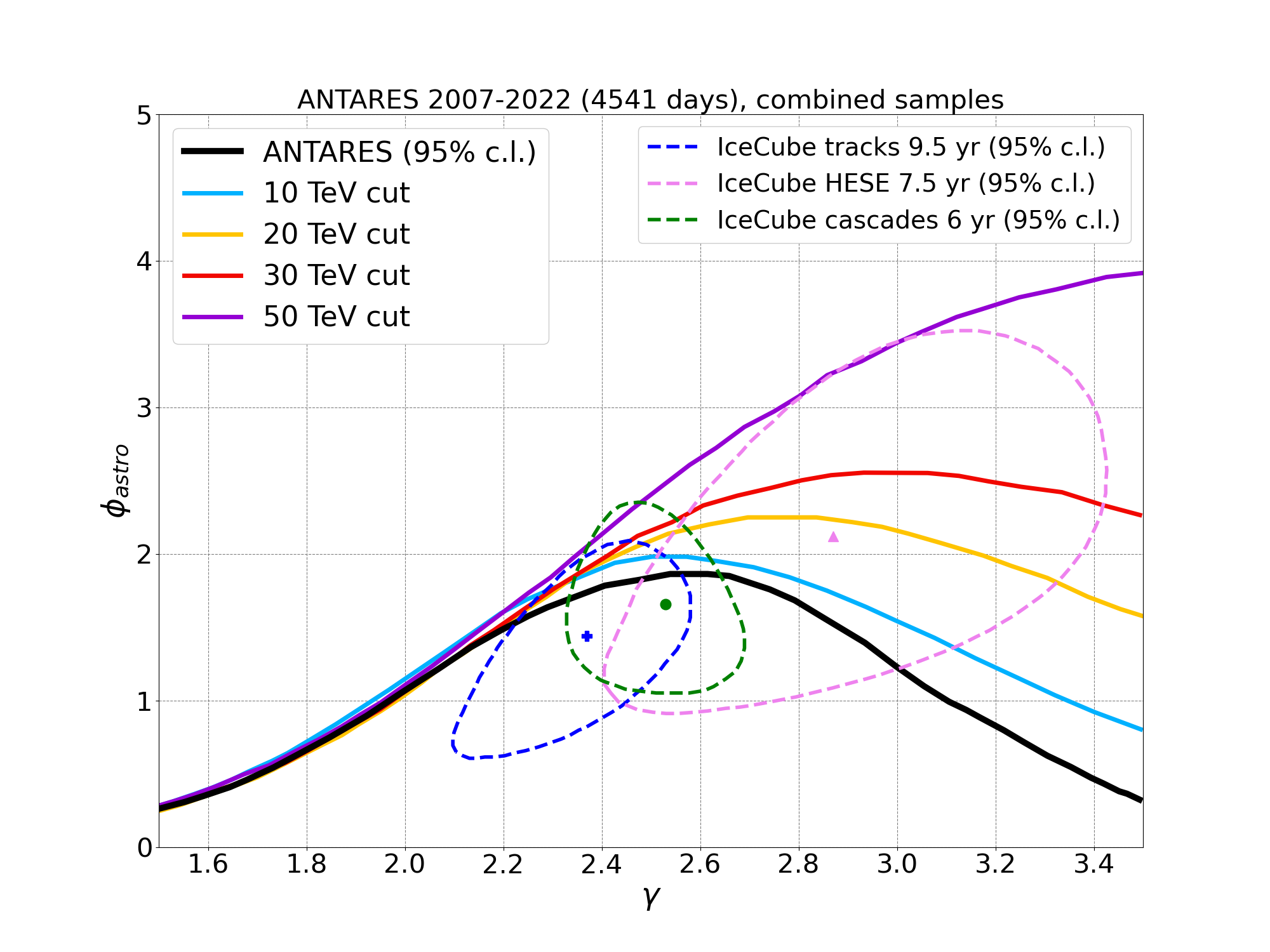}
  \caption{The 95\% posterior probability credible areas obtained from the ANTARES fit assuming the single unbroken power-law hypothesis (black) and adding a low-energy cut in the spectrum from 10 to 50~TeV (coloured lines as in the legend) are compared to the 95\% confidence limit contours from the IceCube HESE (pink), tracks (blue) and cascades (green) samples, shown as dashed lines together with their respective best-fit point.}
  \label{fig:contours_tcut}
\end{figure}

A quantitative estimation can be obtained performing again the fitting procedure, but now assuming this low-threshold cut-off template for the signal, for all channels. The $E_{\nu}^{\textrm{cut}}$ values reported above are tested. The best-fit result remains compatible with the absence of a cosmic signal in all cases. For $E_\nu^{\textrm{cut}} \leq 5$~TeV no difference emerges in the fit results. The 95\% posterior probability credible areas obtained for $E_\nu^{\textrm{cut}}$ values of 10, 20, 30, and 50~TeV are shown in figure~\ref{fig:contours_tcut}, compared with the 95\% confidence level contours from the IceCube HESE, track, and cascade samples. The consequence of the absence of a significant excess of events in the ANTARES dataset is that a single power-law cosmic spectrum described by the HESE best-fit parameters is inside the 95\% ANTARES credible area only if that power law does not extend below 20~TeV, even though these results do not allow to quantitatively state a preference for such cut-off. More complex and less extreme cut-offs than a simple step function could be present (broken power law, log-parabola~\cite{bib:ic_comb}), but given our limited statistics, the eventual outcome of this study assuming different shapes would not yield very different results. 

On a final note, \textit{prompt} atmospheric neutrinos~\cite{bib:prompt_enb, bib:prompt_prosa} originating from the decay of short-lived charmed hadrons in the cosmic ray extensive air showers have not been considered in this work. Upper limits on their contribution have been set by the IceCube Collaboration~\cite{bib:ic_tracks, bib:ic_cascades}, with the best fit for the prompt contribution being always compatible with a null flux. Considering the IceCube constraints, the ANTARES sensitivity to the presence of a prompt neutrino signal is limited. Neglecting the contribution from prompt atmospheric neutrinos does not change any of the obtained results: if a prompt contribution were present, the constraints reported in this work would only become more restrictive so, in the absence of a significant signal, this choice is conservative.

\section{Highest energy events} \label{sec:he_events}

This data analysis selects few outstanding high-energy events both in the track and shower samples. Similarly to the IceCube definition, the signalness $s$ of one individual neutrino event~\cite{bib:icecat} is
\begin{equation}
  s(E^{tr/sh}_{\textrm{reco}}) = \frac{N_{\textrm{signal}}(E^{tr/sh}_{\textrm{reco}})}{N_{\textrm{signal}}(E^{tr/sh}_{\textrm{reco}})+N_{\textrm{background}}(E^{tr/sh}_{\textrm{reco}})}
  \label{eq:signalness}
\end{equation}
where $E^{tr/sh}_{\textrm{reco}}$ is the estimated energy for the track ($tr$) or shower ($sh$) event, and $N_{\textrm{signal}}$ and $N_{\textrm{background}}$ are the expected number of signal and background events at that energy: the former is estimated assuming a signal described by the best-fit flux from the IceCube tracks~\cite{bib:ic_tracks}, while the latter is estimated with the atmospheric flux assumption~\cite{bib:honda}. The signalness of an event is a number in the [0, 1] interval, and signalness values closer to unity tell that the event is more likely of being of cosmic origin. Considering all the samples, 3 events have a signalness value above 0.66 --- so that the neutrino event is at least two times more likely of being of astrophysical than of atmospheric origin. One is a track event (E\"arendil), two are showers (Beren and Luthien). Three additional shower events have signalness between 0.5 and 0.66.

\begin{figure}
  \begin{center}
    \begin{minipage}{0.63\textwidth}
      \begin{center}          E\"arendil \end{center}

      \includegraphics[width=\textwidth]{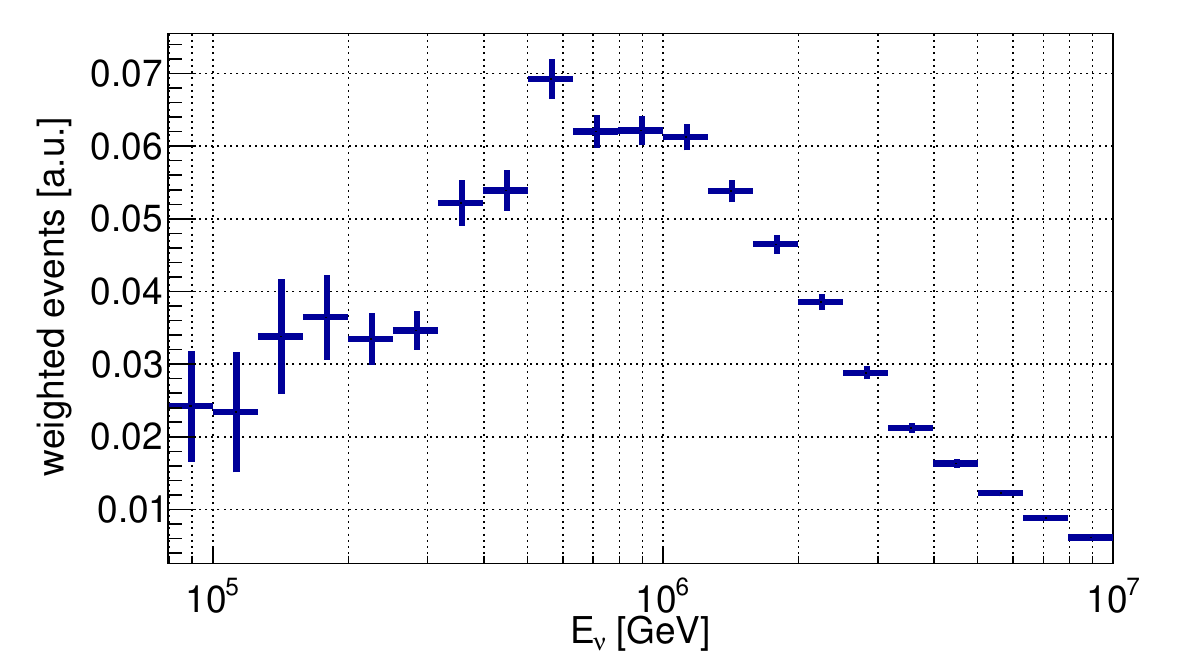}
    \end{minipage}

    \vspace{0.03\textwidth}
    
    \begin{minipage}{0.63\textwidth}
      \begin{center}          Beren \end{center}
      
      \includegraphics[width=\textwidth]{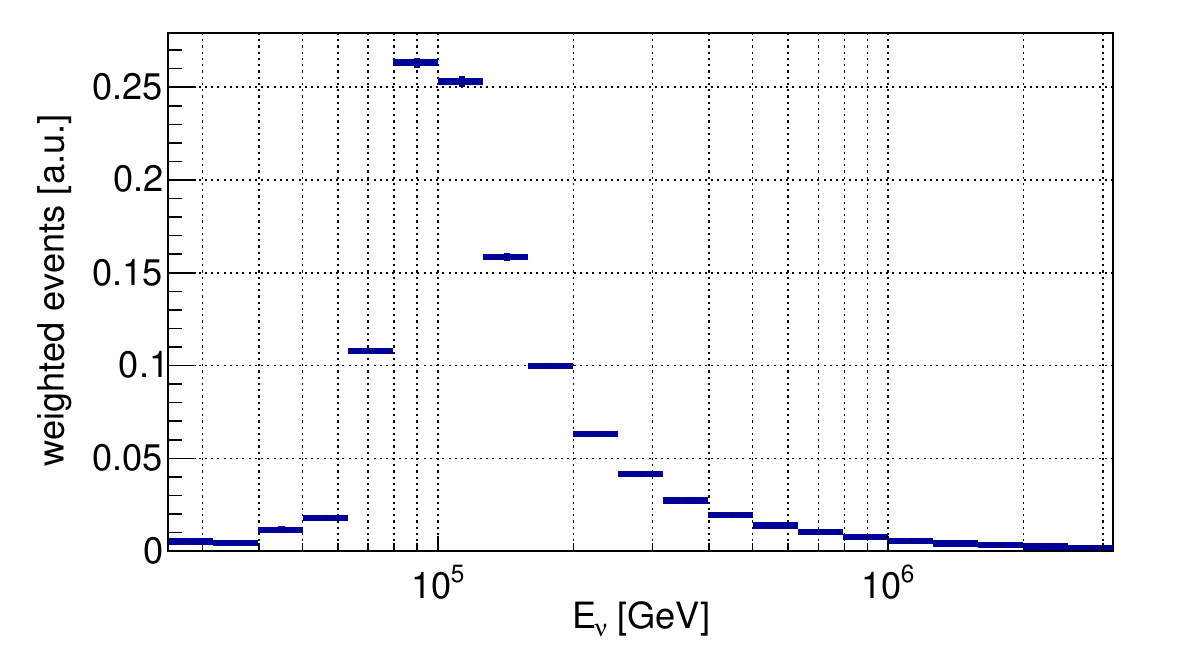}
    \end{minipage}

    \vspace{0.03\textwidth}
    
    \begin{minipage}{0.63\textwidth}
      \begin{center}          Luthien \end{center}
      
      \includegraphics[width=\textwidth]{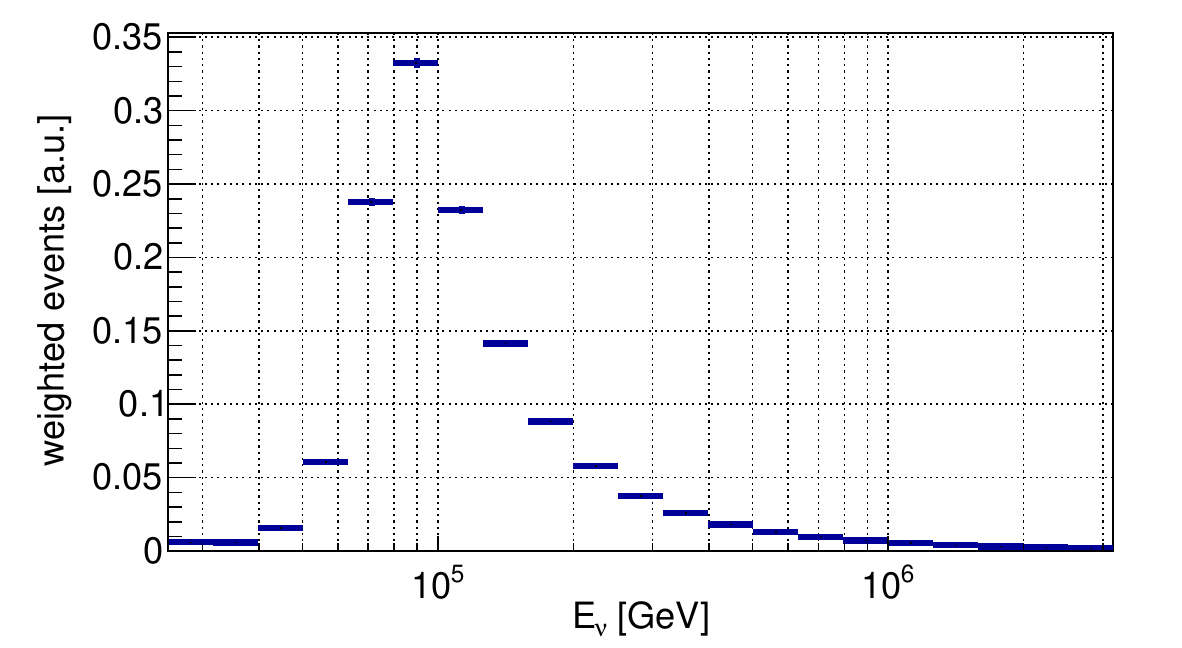}
    \end{minipage}
  \end{center}
  \caption{True neutrino energy ($E_\nu$) distribution for events reconstructed with the same estimated energy as the most energetic track in the sample (E\"arendil), and the two most energetic showers (Beren and Luthien). These distributions are used to estimate the true neutrino energy of the events and the confidence interval in this estimation. Systematic uncertainties are included in these plots.}
  \label{fig:etrue_events}
\end{figure}

In general, the reconstructed energy of the event does not correspond to the parent neutrino energy. The actual neutrino energy associated to the three events can be estimated from simulations. Assuming the same spectra for signal and background events as in the computation of the signalness, the true neutrino energy distributions can be built for events that are reconstructed with the same estimated energy. These distributions are shown in figure \ref{fig:etrue_events}. The median of these distributions is taken as the best estimation of the neutrino energy of each individual event: 700~TeV for the track event, 110 and 95~TeV for the two showers, respectively. The 68\% uncertainty range is estimated from these same distributions, taking the 16\% and 84\% quantiles; systematic uncertainties coming from the limited knowledge of the optical properties of water and of the optical module efficiencies are included in this estimation. All the relevant information on these three events is given in table~\ref{tab:heevents}. No obvious correlation with possible candidate sources has been found in astronomical catalogues~\cite{bib:rfc, bib:4FGL, bib:TeVCat, bib:BZCat, bib:3HSP, bib:3LAC}.

\begin{table}
  {\small
    \begin{center}
      \begin{tabular}{c | c | c c | c c c c} 
        Event name & Type & $E_\nu$ & $E_\nu$ 68\% range & $T$ & ($\delta$, $RA$) & $\beta$ & $s$\\
        & & [TeV] & [TeV] & [MJD] & [deg]  & [deg] & \\ 
        \hline
        E\"arendil & track & 700 & [240, 2300] & 58813.9136016 & (-21.90, 156.38) & 0.31 & 0.66\\
        \hline
        Beren & shower & 110 & [80, 210] & 55562.2854789 & (-82.27, 246.70) & 0.5 & 0.69 \\
        Luthien & shower & 95 & [70, 180]& 56473.3361997 & (-12.82, 190.99) & 2.0 & 0.66\\
      \end{tabular}
    \end{center}
  }
  \caption{Information on the highest energy track (E\"arendil), and the two highest energy showers (Beren and Luthien). The best estimate of the neutrino energy $E_\nu$ and the 68\% energy range are estimated as described in the text. The J2000 equatorial coordinates (declination $\delta$, right ascension $RA$) are shown, together with the time $T$ of occurrence of the event. The estimated angular error $\beta$ from the event reconstruction is reported, as well as the signalness $s$ as defined in equation~\ref{eq:signalness}.}
  \label{tab:heevents}
\end{table}

\section{Conclusions}\label{sec:concl}

The properties of the diffuse cosmic neutrino flux have been investigated using the 15-year dataset of the ANTARES neutrino telescope, collected between 2007 and 2022. Refined data samples have been defined, with a purer neutrino selection, leading to reduced systematic uncertainties with respect to previous works. The distributions of the estimated neutrino energy have been compared to detailed Monte Carlo simulations to search for a high-energy signal of cosmic neutrinos and to study its energy spectrum. The measured distributions for the selected neutrino events are statistically compatible with the background assumptions of only atmospheric neutrinos.

Taking advantage of the large neutrino detection efficiency of ANTARES below 50~TeV, the hypothesis of a low-energy spectral break in the single power-law energy spectrum assumption has been investigated. The hypothesis that a single power-law spectrum with spectral index larger than 2.9 and normalisation at 100~TeV larger than $2\times10^{-18}\textrm{ (GeV cm$^2$ s sr)$^{-1}$}$ extends below 10~TeV is excluded with a 99.7\% Bayesian posterior probability. Such soft-spectra solutions become admissible at a 95\% posterior probability only if a hard low-threshold cut-off is present at least somewhere in the 10~--~30~TeV region. This result is in agreement with the fact that, assuming a power-law extrapolation of soft-spectra fits like the one obtained in the IceCube HESE analysis, the resulting gamma-ray flux from the same hadronic interactions would overshoot the observed extra-galactic gamma-ray background~\cite{bib:stb}. In addition, piece-wise fits of the combined IceCube samples also point in the direction of a possible spectral break for energies around a few tens of TeV.

The ANTARES data taking ended in 2022. In the meanwhile, the construction of the KM3NeT/ARCA neutrino telescope~\cite{bib:km3net} in the Mediterranean Sea has been going on steadily, with first results in the search for the diffuse cosmic signal already being produced~\cite{bib:km3net_diffuse}. The upcoming step will be the combination of the KM3NeT/ARCA data with the 15 years of ANTARES, which will possibly enrich the outcome of this search. Finally, with the increasing size of the KM3NeT/ARCA detector, an additional complementary view on the cosmic neutrino signal will be provided.

\section*{Acknowledgements}

The authors acknowledge the financial support of the funding agencies:
% France:
Centre National de la Recherche Scientifique (CNRS), Commissariat \`a l'\'ener\-gie atomique et aux \'energies alternatives (CEA), Commission Europ\'eenne (FEDER fund and Marie Curie Program), LabEx UnivEarthS (ANR-10-LABX-0023 and ANR-18-IDEX-0001), R\'egion Alsace (contrat CPER), R\'egion Provence-Alpes-C\^ote d'Azur, D\'e\-par\-tement du Var and Ville de La Seyne-sur-Mer, France;
% Germany: 
Bundesministerium f\"ur Bildung und Forschung (BMBF), Germany; 
% Italy
Istituto Nazionale di Fisica Nucleare (INFN), Italy;
% Netherlands
Nederlandse organisatie voor Wetenschappelijk Onderzoek (NWO), the Netherlands;
% Romania
Romanian Ministry of Research, Innovation and Digitalisation (MCID), Romania;
% Spain
MCIN for PID2021-124591NB-C41, -C42, -C43 and PDC2023-145913-I00 funded by MCIN/AEI/10.13039/501100011033 and by “ERDF A way of making Europe”, for ASFAE/2022/014 and ASFAE/2022 /023 with funding from the EU NextGenerationEU (PRTR-C17.I01) and Generalitat Valenciana, for Grant AST22\_6.2 with funding from Consejer\'{\i}a de Universidad, Investigaci\'on e Innovaci\'on and Gobierno de Espa\~na and European Union - NextGenerationEU, for CSIC-INFRA23013 and for CNS2023-144099, Generalitat Valenciana for CIDEGENT/2018/034, /2019/043, /2020/049, /2021/23, for CIDEIG/2023/20 and for GRISOLIAP/2021/192 and EU for MSC/101025085, Spain;
% Marocco
Ministry of Higher Education, Scientific Research and Innovation, Morocco, and the Arab Fund for Economic and Social Development, Kuwait.
% A.O.B.:
We also acknowledge the technical support of Ifremer, AIM and Foselev Marine for the sea operation and the CC-IN2P3 for the computing facilities.

\footnotesize{
  
}

\end{document}